\documentclass[11pt,prd,reprint,twocolumn,notitlepage,nofootinbib,superscriptaddress]{revtex4-1}

\usepackage{float}
\usepackage{amsmath,bm}
\usepackage{amsfonts}
\usepackage{graphicx, adjustbox}
\usepackage{multirow}
\usepackage{aas_macros}
\usepackage[title]{appendix}
\usepackage{xcolor}
\usepackage{hyperref}
\usepackage{orcidlink}
\usepackage{lineno}

\definecolor{mpl_blue}{HTML}{1F77B4}
\definecolor{mpl_orange}{HTML}{FF7F0E}
\definecolor{mpl_green}{HTML}{2CA02C}
\definecolor{mpl_red}{HTML}{D62728}

\newcommand{\Fp}{\ensuremath{\mathcal{F}_p}}
\newcommand{\Fe}{\ensuremath{\mathcal{F}_e}}
\newcommand{\mc}{\ensuremath{\mathcal{M}}}

\newcommand{\shashwat}[1]{\textcolor{mpl_green}{\it{\textbf{shashwat: #1}}} }

\newcommand{\fastfp}{\texttt{fastfp}}

\newcommand{\enterprise}{\texttt{enterprise}}

\newcommand{\entext}{\texttt{enterprise\_extensions}}
\newcommand{\ptmcmc}{\texttt{PTMCMCSampler}}

\date{\today}

\begin{document}

\title{Optimal strategies for
continuous wave detection in pulsar timing arrays: \\Realistic pulsar noise and a gravitational
wave background}

\author{Shashwat C.~Sardesai \orcidlink{0009-0006-5476-3603}}
\affiliation{Center for Gravitation, Cosmology and Astrophysics, University of Wisconsin--Milwaukee, P.O. Box 413, Milwaukee WI, 53201, USA}
\affiliation{Department of Physics and Astronomy, Texas Tech University, Lubbock, TX 79409, USA}

\author{Gabriel E.~Freedman \orcidlink{0000-0001-7624-4616}}
\affiliation{Center for Gravitation, Cosmology and Astrophysics, University of Wisconsin--Milwaukee, P.O. Box 413, Milwaukee WI, 53201, USA}

\author{Sarah J.~Vigeland \orcidlink{0000-0003-4700-9072}}
\affiliation{Center for Gravitation, Cosmology and Astrophysics, University of Wisconsin--Milwaukee, P.O. Box 413, Milwaukee WI, 53201, USA}

\author{Caitlin A. Witt \orcidlink{0000-0002-6020-9274}}
\affiliation{Department of Physics, Wake Forest University, Winston-Salem, NC 27103, USA}

\begin{abstract}
Pulsar timing arrays are sensitive to 
low-frequency gravitational waves (GWs), such as those produced by supermassive binary black holes 
at subparsec separations. 
The incoherent superposition of GWs emitted by a cosmological population 
of these sources produces a gravitational wave background (GWB), 
while some individual sources may be resolvable as deterministic signals 
with slowly varying GW frequencies, which are often referred to as 
``continuous waves'' (CWs). 
The $\mathcal{F}_p$-statistic is a frequentist method of detecting these CWs. In this paper, we study how the presence of pulsar red noise and a GWB affect the \Fp-statistic. 
We compare results when marginalizing over the red noise 
and using the maximum-likelihood values of the red noise, and find little difference between the two. 
We also present results of using the \Fp-statistic to analyze the 
NANOGrav 12.5-year data set, 
where we find no evidence for CWs in agreement with 
the previously published Bayesian results.

\end{abstract}

\maketitle

\section{Introduction}

Pulsar timing arrays (PTAs) use millisecond pulsars to detect the presence of low-frequency gravitational waves (GWs) \cite{1978SvA....22...36S, 1979ApJ...234.1100D, 1990ApJ...361..300F}. When a passing GW passes between the line-of-sight between the Earth and the pulsar, it induces a redshift which results in a delay in the times of arrival of the pulsar signals. By measuring the time-delay (or residuals) $\delta t$, we can detect the presence of GWs. Millisecond pulsars are ideal for this purpose owing to their exceptional rotational stability -- they can be timed to sub-microsecond precision over decades \cite{1997A&A...326..924M, 2009MNRAS.400..951V}
Many astrophysical and cosmological sources are predicted to generate 
low-frequency GWs that can be observed by PTAs, 
including supermassive black hole binaries (SMBHBs) at sub-parsec 
separations ($f \sim 10^{-9} - 10^{-7}$ Hz) \cite{1995ApJ...446..543R, 2003ApJ...583..616J, 2003ApJ...590..691W, 2015MNRAS.451.2417R}. 
Recently, PTA collaborations around the world have published the first 
evidence of nanohertz GWs, which is consistent with predictions for a gravitational wave background (GWB) produced by a population of SMBHBs \cite{2023ApJ...951L...8A, 2023A&A...678A..50E, 2023ApJ...951L...6R, 2023RAA....23g5024X,2025MNRAS.536.1489M}.

If the GWB is coming from a population of SMBHBs, 
we expect that some binaries can be resolved as individual sources \cite{2022ApJ...941..119B}. 
These sources would appear as \textit{continuous waves} (CWs) 
because the GW frequency evolves very slowly compared to the timespan of the data set. 
So far searches of NANOGrav data sets have found no evidence of CWs \citep{NG_11yr_cw, 2023ApJ...951L..28A, 2023ApJ...951L..50A}, 
but we expect that future PTA data sets will detect an individual binary within the next 5-10 years \cite{2022ApJ...941..119B,2025ApJ...988..222G}.

PTAs have primarily performed Bayesian searches for CWs, in which the pulsars' intrinsic noise is simultaneously fit along with GW signals. These searches are computationally expensive owing to the large number of parameters in the model and can take weeks to run. In contrast, frequentist searches are significant faster to perform, although they use simplified signal models. Therefore, frequentist methods complement Bayesian methods and can be used to identify potential signals which can then be followed up with a full Bayesian search.

The $\mathcal{F}$-statistic is a frequentist method to search for CWs 
that was initially developed for the Laser Interferometer Gravitational Wave Observatory (LIGO) \citep{1998PhRvD..58f3001J}. 
The same approach was then applied to the Laser Interferometer Space Antenna (LISA) \citep{2007CQGra..24.5729C} 
and PTAs \cite{2010PhRvD..81j4008S,2012PhRvD..85d4034B}. 
The \Fp- and $\mathcal{F}_e$-statistics, 
introduced in \cite{2012ApJ...756..175E}, 
are frequentist statistics for detecting continuous waves in PTA data. 
The \Fp-statistic was used to analyze the NANOGrav 5-year and 11-year data sets \citep{2014ApJ...794..141A, NG_11yr_cw}, 
but has not been used to analyze subsequent data sets owing to the emergence 
of a common red noise signal. 
In the derivation of the \Fp- and \Fe-statistics, the authors assumed that only white noise and a CW signal were present, and did not consider the presence of either intrinsic red noise or a GWB. 
Recently, work has been done to use the \Fe-statistic to analyze PTA data containing common red noise. 
The European Pulsar Timing Array (EPTA) 
used the \Fe-statistic to search for CWs 
in their second data release 
in addition to a common process \cite{2024A&A...690A.118E}, 
and \citet{2025arXiv250510284F} looked at how 
the presence of a GWB effects the \Fe-statistic.

In this paper, we look at the performance of the 
\Fp-statistic in the presence of pulsar red noise and common red noise. In previous analyses \cite{2023ApJ...951L..28A, 2019ApJ...880..116A, 2014ApJ...794..141A}, we used the maximum likelihood posterior parameters to calculate the \Fp-statistic (MLFP). However, in the case of the optimal statistic, 
a frequentist detection statistic for the GWB \cite{2015PhRvD..91d4048C}, it was found that noise marginalization leads to a more accurate distribution of recovered values \cite{2018PhRvD..98d4003V}. In this paper, we investigate whether marginalizing over the red noise posteriors affects the \Fp-statistic. We find that there is not a significant difference between the mean noise-marginalized \Fp-statistic (NMFP) values and the maximum-likelihood \Fp-statistic (MLFP) values. We use simulations to show that, in the absence of a CW signal, the NMFP values follow the expected chi-squared distribution for frequencies up to where the red noise is modeled, and in 
the presence of a CW the \Fp-statistic can recover 
the signal at the correct frequency.

This paper is organized as follows. In Section \ref{sec:fp}, we discuss our methods and software, including the implementation of the $\mathcal{F}_p$-statistic and its differences from the $\mathcal{F}_e$-statistic as described in \cite{2012ApJ...756..175E}. 
We tested the \Fp-statistic on simulated PTA data sets 
to study its performance in the presence of intrinsic pulsar red noise, 
common red noise, and a continuous wave; we present the results of 
our analyses on simulated data sets in Section \ref{sec: results}. 
In Section \ref{sec: 12.5yr results}, we present the noise-marginalized results for the 
NANOGrav 12.5-year data set \citep{2021ApJS..252....4A}, 
and compare these results to the results of Bayesian searches for individual binaries published in \cite{2023ApJ...951L..28A}. 
We give a brief summary of our work and our conclusions in Section \ref{sec: conclusions}.

\section{Methods and software} \label{sec:fp}

In this section, we review the signal model for a CW in PTA data. 
We assume the source is a circular binary evolving only due to GW emission. 
We also review the \Fp-statistic, a frequentist statistic 
for detecting CWs. More details can be found in \cite{2012ApJ...756..175E}.

\subsection{Signal model}

GWs induce correlated changes in the pulse times arrival times 
for millisecond pulsars. 
The differences between the expected and observed pulse times of arrival are the \textit{residuals}. 
We can write the residuals for each pulsar $\delta t$ as \cite{Lynch:2015iua}
\begin{equation}
	\delta t = M \epsilon + Fa + n + s \,, \label{eq:residuals}
\end{equation}
where $M$ is the design matrix, which describes the linearized timing model, 
$\epsilon$ is a vector of the timing model parameter offsets, 
$Fa$ describes red noise, 
$n$ is a vector describing white noise, 
and $s$ is a vector of the residuals induced by a CW.

The response of a pulsar 
with sky position $(\theta,\phi)$ to a GW source is given by 
the antenna pattern functions \cite{2012ApJ...756..175E}

\begin{eqnarray}
	F^+(\hat{\Omega}) &=& \frac{1}{2} \frac{(\hat{m} \cdot \hat{p})^2 - (\hat{n} \cdot \hat{p})^2}{1+\hat{\Omega} \cdot \hat{p}} \,, \\
	F^\times(\hat{\Omega}) &=& \frac{(\hat{m} \cdot \hat{p}) (\hat{n} \cdot \hat{p})}{1+\hat{\Omega} \cdot \hat{p}} \,,
\end{eqnarray}
where 
\begin{eqnarray}
    	\hat{m} &=& -\sin\phi \; \hat{x} + \cos\phi \; \hat{y} \,, \\
	\hat{n} &=& -\cos\theta \cos\phi \; \hat{x} - \cos\theta \sin\phi \; \hat{y} + \sin\theta \; \hat{z} \,,
\end{eqnarray}
and $\hat{\Omega}$ is a unit vector pointing from the 
GW source to the Solar System barycenter. 
The induced residuals can be written
\begin{equation}
	s(t, \hat{\Omega}) = F^+(\hat{\Omega}) \; \Delta s_+(t) + F^\times(\hat{\Omega}) \; \Delta s_\times(t) \,,
\end{equation}
where 
$\Delta s_{+,\times}$ is the difference between the signal induced at the Earth and at the pulsar 
(the so-called ``Earth term'' and ``pulsar term''), 
\begin{equation}
	\Delta s_{+,\times}(t) = s_{+,\times}(t_p) - s_{+,\times}(t) \,.
\end{equation}
The time at which the GW passes the Solar System barycenter $t$ 
and the time at which it passes the pulsar $t_p$ are related by
\begin{equation}
	t_p = t - L (1 + \hat{\Omega} \cdot \hat{p}) \,,
	\label{eq:pulsar_time}
\end{equation}
where $L$ is the distance to the pulsar and $\hat{p}$ is a unit vector pointing from the Earth to the pulsar.

For a circular binary the signal is given by \citep{wahlquist1987,lwk+2011,cc2010} 
\begin{eqnarray}
	s_+(t) &=& \frac{\mc^{5/3}}{d_L \, \omega(t)^{1/3}} \left[ -\sin 2(\Phi(t)-\Phi_0) \, \left(1+\cos^2i\right) \, \cos2\psi \right. \nonumber \\
			&& \left. - 2 \cos 2(\Phi(t)-\Phi_0) \, \cos i \, \sin 2\psi \right] \,, \label{eq:signal1} \\
	s_\times(t) &=& \frac{\mc^{5/3}}{d_L \, \omega(t)^{1/3}} \left[ -\sin 2(\Phi(t)-\Phi_0) \, \left(1+\cos^2i\right) \, \sin2\psi \right. \nonumber \\
			&& \left. + 2 \cos2(\Phi(t)-\Phi_0) \, \cos i \, \cos 2\psi \right] \,, \label{eq:signal2}
\end{eqnarray}
where $\omega$ is the orbital frequency, $i$ is the inclination angle of the SMBHB, $\psi$ is the GW polarization angle, 
$d_L$ is the luminosity distance to the source, 
and $\mc \equiv (m_1 m_2)^{3/5}/(m_1+m_2)^{1/5}$ 
is a combination of the black hole masses $m_1$ and $m_2$ 
called the ``chirp mass.'' 
Note that the variables $\mc$ and $\omega$ are the observed redshifted values,
which are related to the rest-frame values $\mc_r$ and $\omega_r$ according to $\mc_r = \mc/(1+z)$ and $\omega_r = \omega(1+z)$. 
Currently PTAs are only sensitive to sources in the local Universe for which $(1+z) \approx 1$.

As the binary emits GWs, 
the orbital frequency of the binary evolves according to
\begin{equation}
    \frac{d\omega}{dt} = \frac{96}{5} \mc^{5/3} \omega^{11/3} \,.
\end{equation}
For typical SMBHBs emitting in the PTA band, 
$d\omega/dt$ is not significant over the timescale of 
the PTA data set ($\sim20$ yrs), 
but can be over the light-travel time between 
the pulsars and the Earth ($\sim1000$ yrs). 
Then the orbital frequency at the pulsar $\omega_p \equiv \omega(t_p)$ 
is related to the orbital frequency at the Earth $\omega_0 \equiv \omega(t)$ 
according to
\begin{equation}
    \omega_p = \omega_0 \left[ 1 + \frac{256}{5} \mc^{5/3} \omega_0^{8/3} L (1 + \hat{\Omega} \cdot \hat{p}) \right]^{-3/8} \,.
\end{equation}
For slowly evolving sources, we can approximate this as 
\begin{equation}
    \omega_p \approx \omega_0 \left[ 1 - \frac{96}{5} \mc^{5/3} \omega_0^{8/3} L (1 + \hat{\Omega} \cdot \hat{p}) \right] \,.
\end{equation}

\subsection{$\mathcal{F}_p$-statistic} \label{sec: fp stat}

The likelihood of a continuous wave being present is given by 

\begin{equation}
    p(\delta t|s) = \frac{\exp[-\frac{1}{2}(\delta t - s)^T C^{-1}(\delta t - s)]}{\sqrt{\mathrm{det}[2 \pi C]}} ,
\end{equation}

\noindent where $C$ is the total covariance matrix of the PTA. For signals uncorrelated between pulsars the covariance matrix becomes a block diagonal matrix of the individual pulsar auto-covariances, i.e., $C = \mathrm{diag}(\Sigma_1, \Sigma_2, \cdots \Sigma_N)$, where $\Sigma_\alpha$ is

\begin{equation} \label{eq: covariance}
    \Sigma_\alpha = N_\alpha + T_\alpha^T B_\alpha T_\alpha,
\end{equation}

\noindent where $T$ is a block matrix containing the design matrix $M$ and the Fourier matrix $F$, $N$ is the white noise covariance matrix, and $B$ is a block diagonal matrix containing the covariances of the priors of parameters $\epsilon$ and $a$. An important note is that in the presence of a common correlated signal, the total covariance matrix $C$ is now a dense block matrix with pulsar auto-covariances $\Sigma_{a}$ in the diagonal terms and cross-covariances $\Sigma_{ab}$, defined by 

\begin{equation} \label{eq: cross covariance}
    \Sigma_{\alpha \beta} = \Gamma_{\alpha \beta} F \phi_\mathrm{GWB} F^T ,
\end{equation}

\noindent where $F$ is the Fourier matrix, $\phi_\mathrm{GWB}$ is the power spectral density of the correlated background, and $\Gamma_{ab}$ is the Hellings-Downs coefficient between pulsars $a$ and $b$ \cite{1983ApJ...265L..39H}. In the case of a common uncorrelated red noise process (CURN), the $T^T B T$ matrix contains intrinsic red noise, and a common red noise process for all pulsars in Eqn. \ref{eq: covariance}. In the case of a correlated common red noise process (the GWB), the total covariance matrix C now contains off diagonal terms that follow Eqn. \ref{eq: cross covariance}.

We start with the log-likelihood ratio of a continuous signal $s$ being present within our residuals $\delta t$ to no signal,

\begin{eqnarray}
    \ln{\Lambda} &=& \ln \frac{p(\delta t | s)}{p(\delta t | 0)} , \\
    \label{eq: chi square}
    \ln{\Lambda} &=& \sum_\alpha (\delta t_\alpha|\Tilde{s}_\alpha) - \frac{1}{2}(\Tilde{s}_\alpha|\Tilde{s}_\alpha) .
\end{eqnarray}

\noindent Here, the terms in the parentheses are the inner product with the inverse covariance matrix, i.e., $(x|y) = x\Sigma^{-1}y$, where $\Sigma$ is the pulsar auto-covariance matrix.

The signal $s$ from a continuous wave source is represented as the sum of sines and cosines for the $\alpha$ pulsars\footnote{Here we are only considering circular SMBBHs. For a discussion of frequentist searches for eccentric SMBBHs, see \cite{Taylor_2016}.}

\begin{equation}
    s_{\alpha} (t, \hat{\Omega}) = \sum_{i=1}^2 b_{i\alpha} (\zeta, \iota, \psi, \Phi_0, \phi_{\alpha}, \theta, \phi) B^i_{\alpha} (t, \omega_0) ,
\end{equation}

\noindent where $\zeta = \mathcal{M}^{5/3}/d_L$ is the amplitude, $\iota$ is the inclination, $\psi$ is the polarization, $\Phi_0$ is the gravitational wave initial phase, $\phi_{\alpha} = \omega L_{\alpha} (1+ \hat{\Omega}\cdot \hat{p}) + \Phi_0$ is the pulsar dependent phase, and $\theta, \phi$ are the sky position for the source. The only intrinsic parameter is $\omega_0$. There are two terms (sine and cosine) for the amplitudes $b_{i \alpha}$ and the basis functions $B^i_{\alpha}$. 

The amplitudes $b_{1\alpha}$ and $b_{2\alpha}$ are given by

\begin{widetext}
\begin{eqnarray}
    b_{1\alpha} &=& \zeta \left[ (1+\cos^2{\iota})(F_{\alpha}^+ \cos{2\psi} + F_{\alpha}^{\times} \sin{2\psi})(\cos{\Phi_0} - \cos{\phi_{\alpha}}) \right. \nonumber \\ 
    && \left. + 2\cos{\iota} (F_{\alpha}^+ \sin{2\psi} -F_{\alpha}^{\times} \cos{2\psi}) (\sin{\Phi_0} - \sin{\phi_{\alpha}})\right] ,\\
    b_{2\alpha} &=& -\zeta \left[ (1+\cos^2{\iota})(F_{\alpha}^+ \cos{2\psi} + F_{\alpha}^{\times} \sin{2\psi})(\sin{\Phi_0} - \sin{\phi_{\alpha}}) \right. \nonumber \\ 
    && \left. - 2\cos{\iota} (F_{\alpha}^+ \sin{2\psi} -F_{\alpha}^{\times} \cos{2\psi}) (\cos{\Phi_0} - \cos{\phi_{\alpha}})\right] ,
\end{eqnarray}
\end{widetext}

\noindent and the basis functions are simply,

\begin{eqnarray}
    B^1_{\alpha} (t) &=& \frac{1}{\omega_0^{1/3}} \sin{2 \omega_0 t} ,\\
    B^2_{\alpha} (t) &=& \frac{1}{\omega_0^{1/3}} \cos{2 \omega_0 t} .
\end{eqnarray}

The log-likelihood then becomes

\begin{equation}
    \ln{\Lambda} = \sum_{\alpha=1}^M \left[ b_{i\alpha}(r_{\alpha}|B^i_{\alpha}) - \frac{1}{2} (B^i_{\alpha}|B^j_{\alpha}) b_{i\alpha} b_{j\alpha} \right] .
\end{equation}

\noindent Here, the indices $i,j$ are for the sine and cosine terms and $M$ is the total number of pulsars. We can represent the inner products as matrices $P^i_{\alpha} = (r_{\alpha}|B^i_{\alpha})$ and $Q^{ij}_{\alpha} = (B^i_{\alpha}|B^j_{\alpha})$, and maximize the log-likelihood to get the relationship $b_{i\alpha} = Q^{\alpha}_{ij} P^{j}_{\alpha}$, where $Q^{\alpha}_{ij} = (Q^{ij}_{\alpha})^{-1}$. The matrices $P_\alpha^i$ has dimensions of $N_{\mathrm{TOA}, \alpha} \times 2$, and $Q^{ij}_\alpha$ is a $2 \times 2$ matrix.

Thus we get the maximized log-likelihood and the $\mathcal{F}_p$ statistic, 

\begin{eqnarray}
    \ln{\Lambda} &=& \frac{1}{2} \sum_{\alpha=1}^M P^{i}_{\alpha} Q^{\alpha}_{ij} P^{j}_{\alpha} ,\\
    2\mathcal{F}_p &=& \sum_{\alpha=1}^M P^{i}_{\alpha} Q^{\alpha}_{ij} P^{j}_{\alpha} .
\end{eqnarray}

Here, the distribution of the 2\Fp-statistic a chi-squared with degrees of freedom $2N$, where $N$ is the number of pulsars, and a non centrality of $\rho^2 = (\Tilde{s}|\Tilde{s})$, in the case of a continuous wave, where $\rho$ is the signal-to-noise ratio (S/N) of the continuous wave. Each frequency bin will follow the standard chi-squared distribution, except for bins with a continuous wave, which will follow a non-central chi squared distribution. Since there are 6 independent extrinsic variables per pulsar, and $2N$ equations, we need a minimum of 6 pulsars to solve for all variables \citep{2012ApJ...756..175E}.

\begin{figure}[!t]
    \centering
    \includegraphics[width=0.75\linewidth]{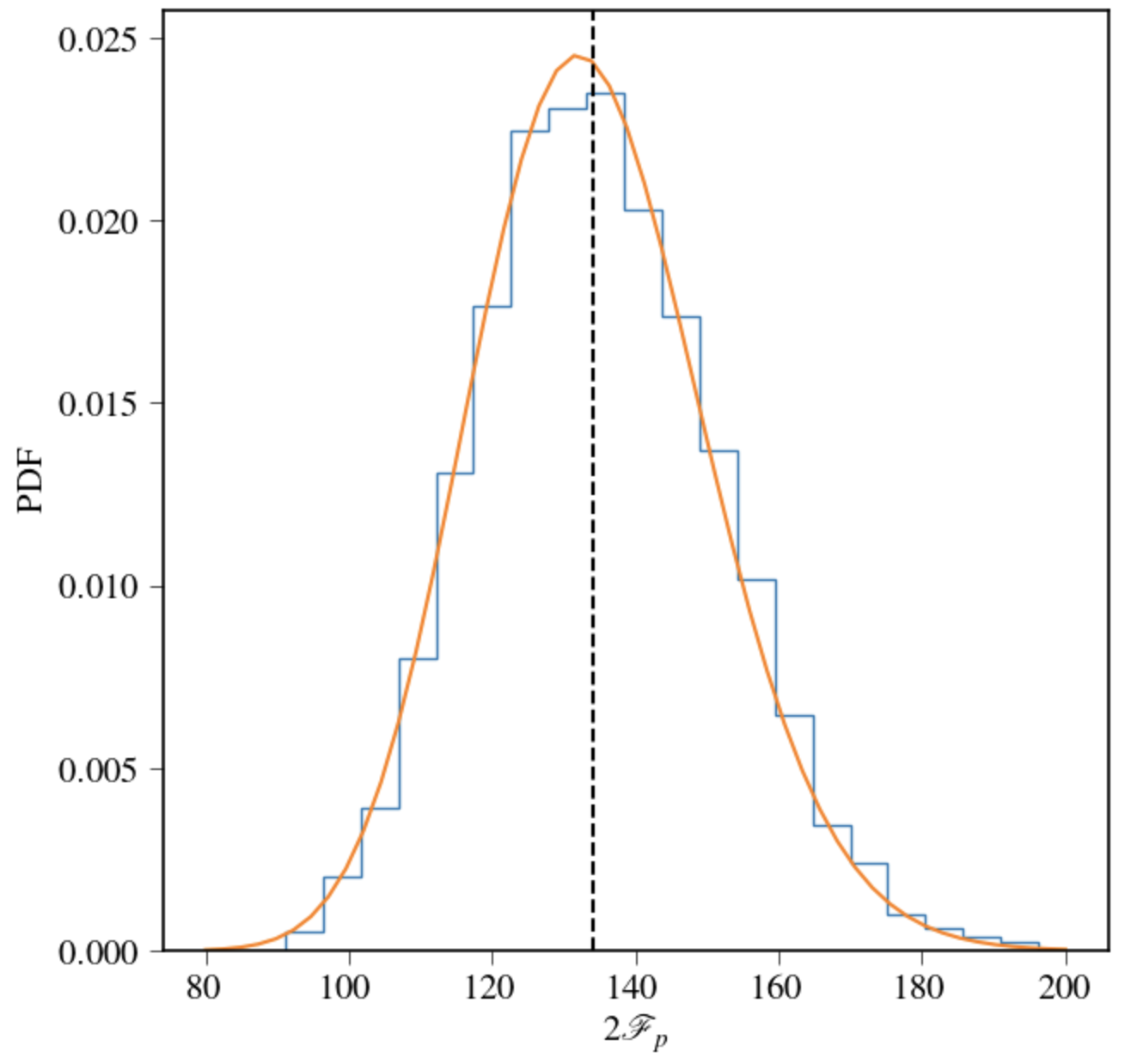}
    \caption[White noise simulations for $\mathcal{F}_p$ statistic]{The blue histogram shows the $2 \mathcal{F}_p$ values for 10 simulations of the 15yr dataset with only white noise injected, marginalized across 150 frequency bins. The orange curve is the expected chi squared distribution with zero non-centrality and the black dashed line is the expected degrees of freedom.}
    \label{fig:wn fp}
\end{figure}

For our simulations based on the NANOGrav 15-year data set, 
there are 67 pulsars, so the expected distribution of the \Fp-statistic 
if there is only white noise present is a chi-squared distribution 
with 134 degrees of freedom, as shown in Fig. \ref{fig:wn fp}. To determine the presence of a continuous wave, we compute the FAP by comparing the measured value of the \Fp-statistic to the null distribution
\citep{2012ApJ...756..175E, 1998PhRvD..58f3001J, 2000PhRvD..61f2001J}. 
The probability distributions for when the signal is absent ($p_0$) and present ($p_1$), and intrinsic parameters are known are

\begin{eqnarray}
    p_0(\mathcal{F}) &=& \frac{\mathcal{F}^{n/2 -1}}{(n/2 -1)!} \exp{(-\mathcal{F})} ,\\
    p_1(\mathcal{F}, \rho) &=& \frac{(2\mathcal{F})^{(n/4-1/2)}}{\rho^{(n/2-1)}} I_{n/2-1}(\rho \sqrt{2\mathcal{F}}) \nonumber \\
    && \times \exp{(-\mathcal{F}-\frac{1}{2}\rho^2)} \,
\end{eqnarray}

\noindent where $n$ is the degrees of freedom ($n=2M$), $\rho$ is the signal-to-noise ratio, and $I_x$ is the modified Bessel function of the first kind. 
The probability that $\mathcal{F}$ exceeds a threshold $\mathcal{F}_0$ when no signal is present is the FAP,
\begin{equation}
    P_F(\mathcal{F}_0) = \int_{\mathcal{F}_0}^\infty p_0(\mathcal{F}) d\mathcal{F} = \exp{(-\mathcal{F}_0)} \sum_{k=0}^{n/2-1} \frac{\mathcal{F}_0^k}{k!} \,.
\end{equation}

\noindent Since we do not know the intrinsic parameters we must adjust the FAP values to reflect the number of independent cells $N_c$, as shown in \cite{2012ApJ...756..175E, 2000PhRvD..61f2001J}

\begin{equation} \label{eq:fp fap}
    P^T_F(\mathcal{F}_0) = 1 - \left[ 1- P_F(\mathcal{F}_0) \right]^{N_c} ,
\end{equation}

\noindent where in our analysis, $N_c$ is the number of frequency bins over which we calculate the $\mathcal{F}_p$ value. This value gives us the probability the $\mathcal{F}$ exceeds $\mathcal{F}_0$ in one or more bins.

The \Fp-statistic is an incoherent search which marginalizes over the sky location of the source. In this respect, it is analogous to Bayesian all-sky searches. 
In contrast, the \Fe-statistic is a coherent search, 
and the sky location of the source and GW frequency are free parameters. In this paper, we only discuss the \Fp-statistic; other papers have examined the \Fe-statistic in the presence of pulsar red noise and common red noise \cite{2024A&A...690A.118E, 2025arXiv250510284F}.

\subsection{Software}

All of the results below utilize our new package \fastfp\footnote{https://github.com/gabefreedman/fastfp} which is publicly available on Github. Both the original $\mathcal{F}_{p}$-statistic as well as the NMFP described above are implemented in our code. It is programmed entirely using \texttt{JAX}\citep{jax2018github}, leading to roughly a $10\times$ increase in speed doing a direct comparison for the base $\mathcal{F}_{p}$ calculation against the previously used code. It can also be natively run using GPU resources, providing an additional decrease in overall runtime.

Additionally we use the standard suite of PTA analysis software to carry our simulated data studies and Bayesian analyses. We use the \texttt{libstempo} package to create the three different configurations of simulated datasets. 
We use \enterprise\ \cite{enterprise} and \entext\ \cite{ent_ext} to set up the PTA model and calculate the likelihood, and we use \ptmcmc\ \cite{justin_ellis_2017_1037579} to perform the MCMC sampling.

\section{Results: Simulated data sets} \label{sec: results}

We use simulated data sets to test the performance 
of the $\Fp$-statistic in the presence of pulsar noise, 
common uncorrelated red noise (CURN), 
a gravitational wave background (GWB), 
and a CW signal. 
We base our data sets on the NANOGrav 15-year data set \cite{2023ApJ...951L...9A}: 
we simulate 67 pulsars observed for up to 15.9 years, 
with sky positions and timing model parameters taken from the NANOGrav 15-year data set. 
We include white noise and intrinsic red noise in each pulsar: 
the injected red noise parameters are given in Table \ref{tab:rn params fp}.
For the simulations including CURN or a GWB, we model the power spectrum 
as a power-law with amplitude and spectral index given in Table~\ref{tab:rn params fp}. 

For the CW injection we use the most sensitive sky position as given by \cite{2023ApJ...951L..50A} with parameters given in Table \ref{tab:cw inj params}.
We considered two CW injections, one at 8 nHz and one at 16 nHz, 
with S/N of 5.4 and 9.1, respectively for cases with a medium and high S/N. The signal to noise ratio can be calculated using Eq.~\eqref{eq: chi square}:

\begin{equation}
    S/N = \sqrt{\sum_\alpha (\tilde{s}_\alpha|\tilde{s}_\alpha)} .
\end{equation}

In the presence of a CW, the non-centrality parameter should be $\rho^2$, 
where $\rho$ is the S/N of the signal.
For data sets containing red noise, 
we computed the \Fp-statistic in two ways: 
using the maximum-likelihood noise parameters, 
and marginalizing over the red noise using the red noise posteriors 
from a Bayesian analysis. 
This type of noise-marginalization 
has been done in frequentist searches for the GWB, 
and more accurately recovers the amplitude and significance of the GWB 
than using maximum-likelihood noise parameters \cite{2018PhRvD..98d4003V,2023PhRvD.108l3007V}. 
However, in our analysis we found little difference between these two methods 
in the \Fp-statistic results.

\subsection{White noise and intrinsic red noise}

\begin{figure}[b!]
    \centering
    \includegraphics[width=\columnwidth]{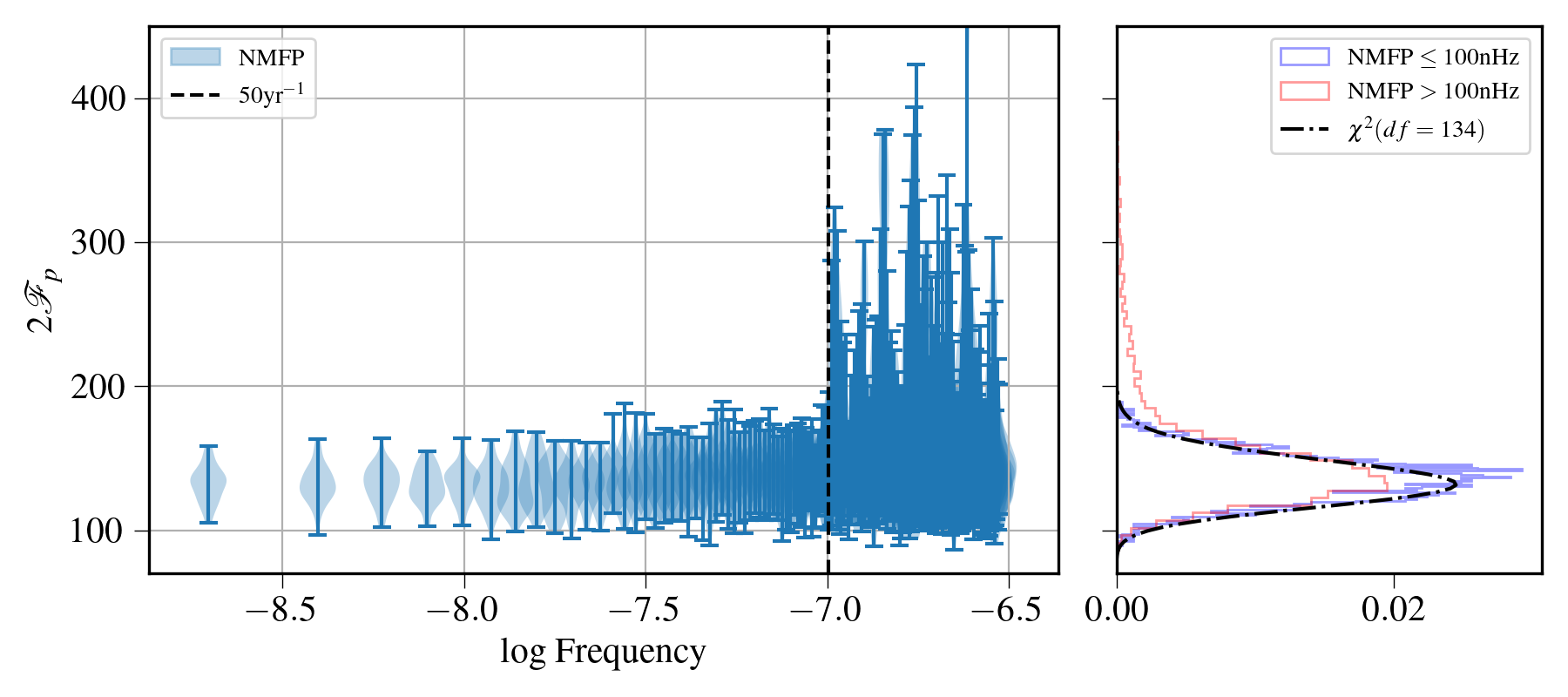}
    \caption{(Left) The 2\Fp\ value for simulations with white noise and intrinsic red noise for the NMFP values. Both the Bayesian and the frequentist analysis were performed for 50 red noise frequency components (black dashed line), beyond which the \Fp value becomes large and unreliable. (Right) The histograms showing the \Fp\ values below 100 nHz ($50 \mathrm{yr}^{-1}$) \Fp\ in light blue, and the $\mathcal{F}_p$ values above 100 nHz in dark blue.
    }
    \label{fig:all 150}
\end{figure}

We considered data sets with both white noise and intrinsic red noise. 
Figure~\ref{fig:all 150} shows the recovered values of the \Fp-statistic 
between 2-300 nHz. 
The recovered values follow the expected chi-squared distribution 
below 100 nHz; 
however, above that the recovered values no longer follow the expected distribution. 
This transition occurs at the frequency 
where we truncate the model for the pulsar's intrinsic red noise, 
demonstrating the importance of modeling pulsar noise 
when searching for CWs. For the remainder of the paper, 
we limit the \Fp-statistic analyses to below 100 nHz. 
In Appendix \ref{p-p plots} we show the recovery of both the NMFP and MLFP across all simulations for each of the modeled frequencies.

\subsection{White noise, intrinsic red noise, and CURN}

\begin{figure}[b!]
    \centering
    \includegraphics[width=\linewidth]{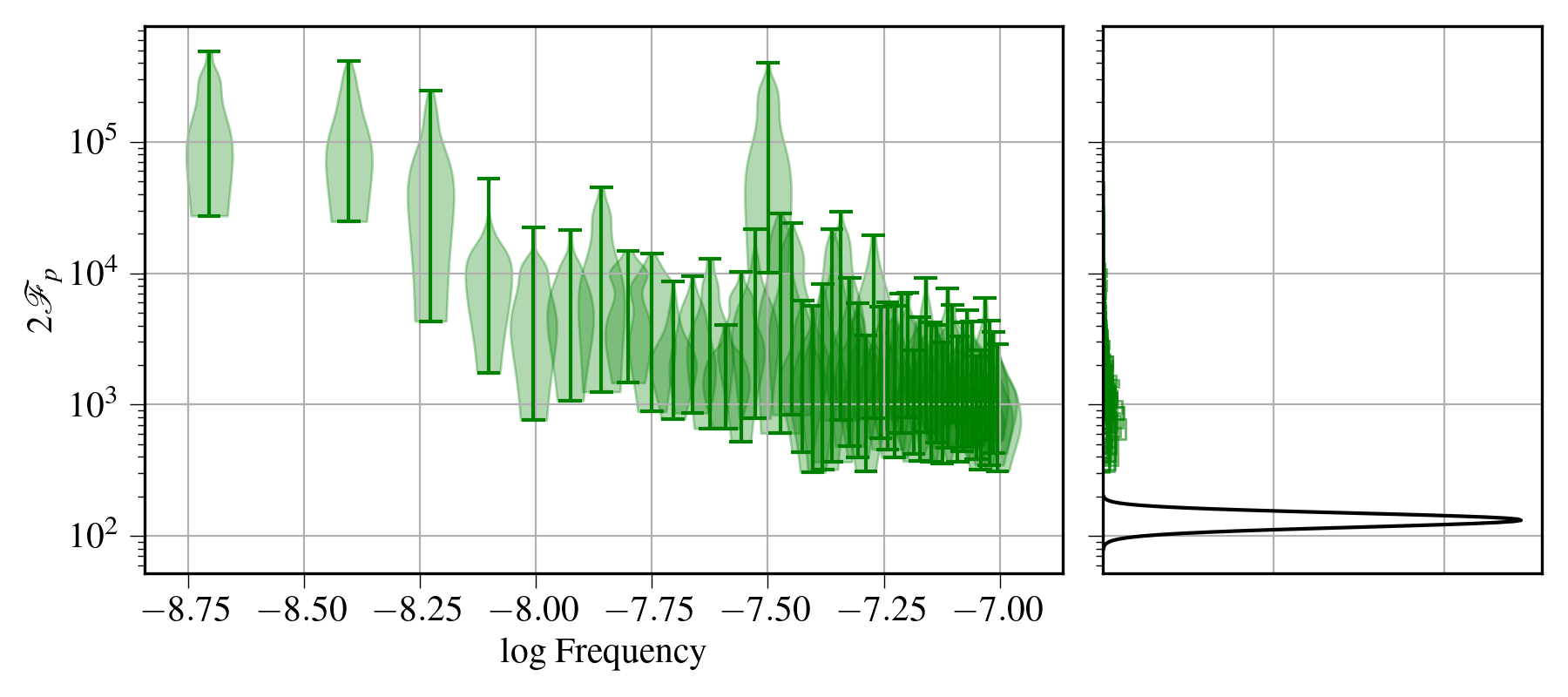} \\
    \includegraphics[width=\linewidth]{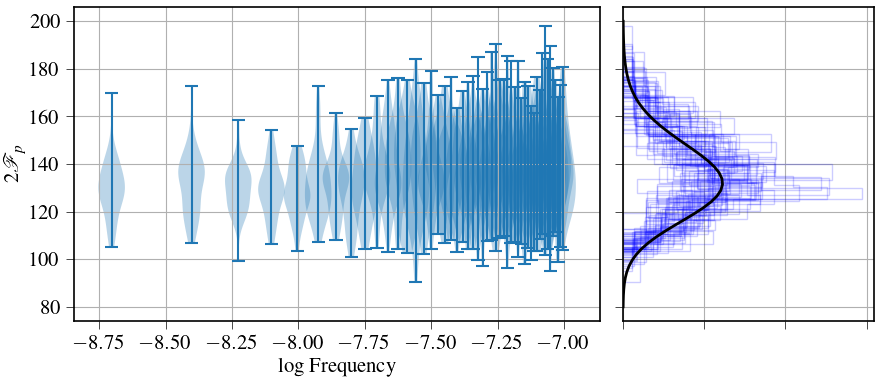} \\
    \includegraphics[width=\linewidth]{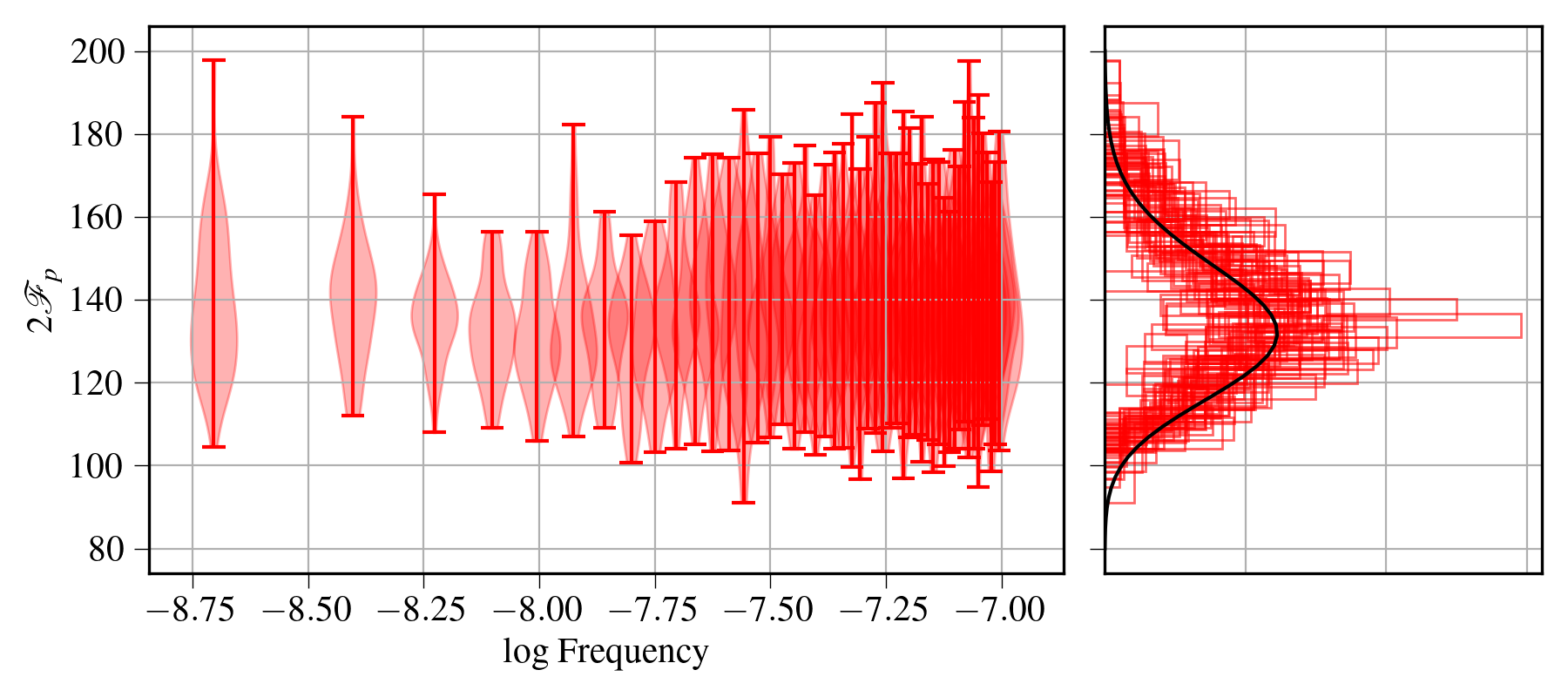}
    \caption[Violin plots for WN+IRN+CURN simulations]{TOP:  (Left) The violin plot and histograms for \Fp\ values when we do not include the CURN model. We see that at all frequencies the \Fp\ values are greatly overestimated. (Right) The histograms for the $2\mathcal{F}_p$ values as compared to the expected central chi-squared distribution with 134 degrees of freedom.  MIDDLE: The violin plots for the mean NMFP values of 50 simulations with white noise, intrinsic red noise, and a common process as a function of frequency. BOTTOM: The figure shows the violin plots for the NMFP values when the common process is kept fixed to the maximum likelihood values and only intrinsic red noise parameters are marginalized over. At low values, the \Fp\ values recover a slightly higher mean and variance due to the common process not being modeled correctly.}
    \label{fig:curn fp}
\end{figure}

Next we applied the \Fp-statistic to data sets containing CURN in addition 
to pulsar white noise and intrinsic red noise. 
When we neglected to model the CURN, 
we found very high values of 2\Fp\ at the lowest frequencies, indicating confusion between CURN and a CW, 
but at higher frequencies 2\Fp\ followed a chi-squared distribution 
with non-centrality parameter of zero. 
This is because the CURN signal has a steep power spectrum, and so it only affects the \Fp-statistic analysis at low frequencies.

When we included the CURN in our model, as described in Sec \ref{sec: fp stat}, 
we again found that 2\Fp\ followed a chi-squared distribution 
with non-centrality parameter of zero, indicating no detection of a CW. 
This is consistent with the findings of \cite{2025arXiv250510284F}, where it was found that if the background is modeled as noise, it can be distinguished from the continuous wave. In the top panel (blue) of Fig.~\ref{fig:curn fp} we see that the means of the NMFP values are in agreement the expected chi-squared distribution across all 50 frequencies when we marginalize over the intrinsic red noise as well as the CURN parameters. In the middle panel (red) we show the recovery when the CURN parameters are fixed to the maximum likelihood posteriors, and only marginalize the intrinsic red noise parameters. 

We find good agreement between the \Fp\ values when we marginalize over the CURN parameters or use the maximum likelihood CURN parameters, although there is larger variance in the \Fp\ values when marginalizing over CURN at very low frequencies.

\subsection{White noise + Intrinsic Red Noise + HD simulations}

\begin{figure}[t!]
    \centering
    \includegraphics[width=\linewidth]{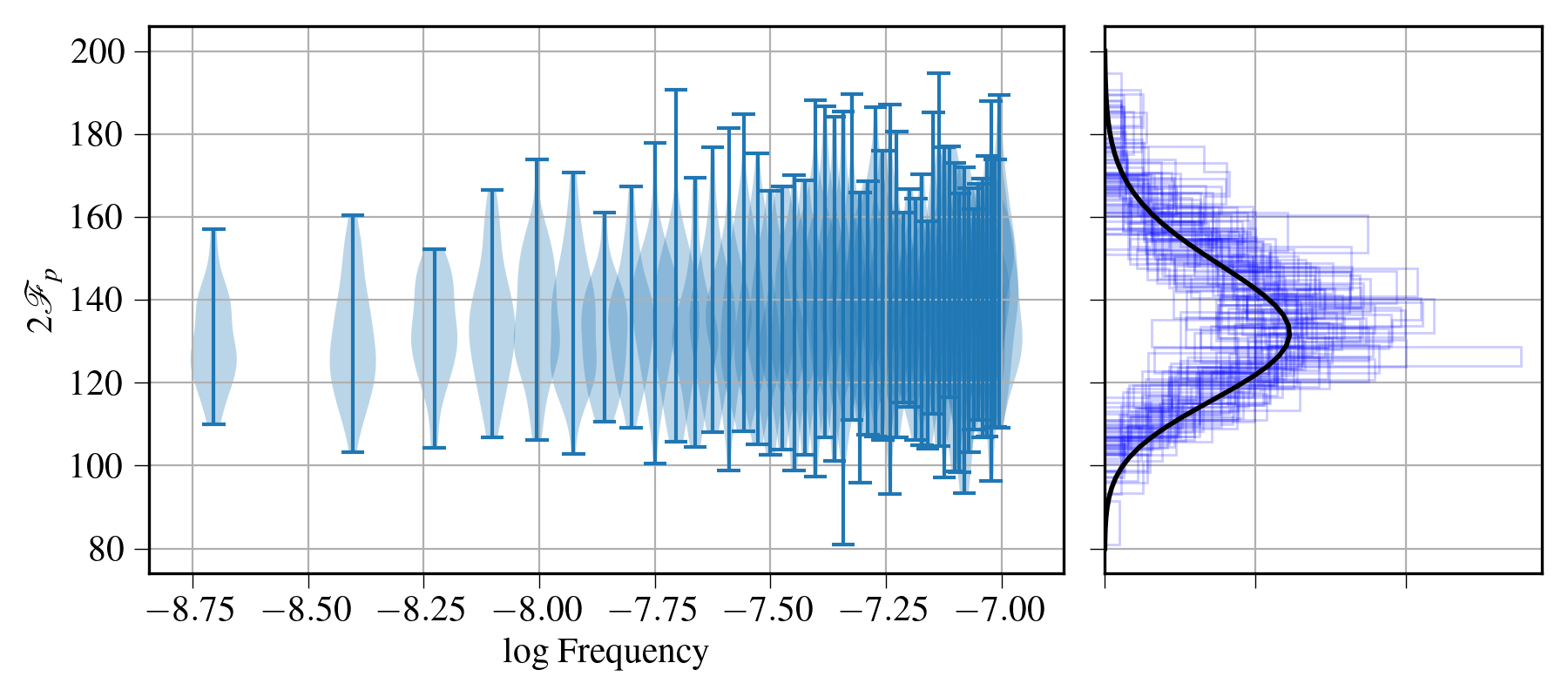} \\
    \includegraphics[width=\linewidth]{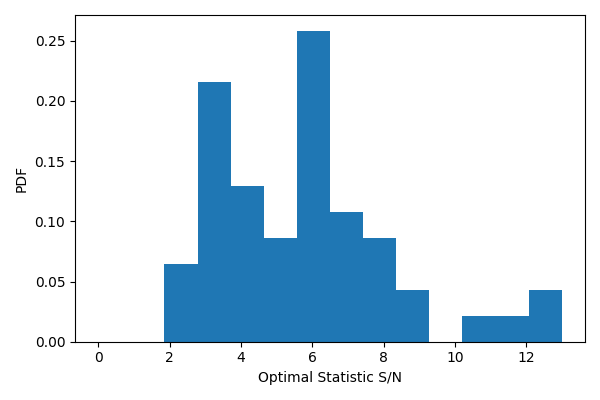}
    \caption{TOP: The violin plots of the NMFP values for simulations with white noise, intrinsic red noise, and a common correlated background (left) and a plot of the histograms of \Fp\ values for each frequency (right). We see that this figure is nearly identical to Fig. \ref{fig:curn fp}. BOTTOM: The S/N of the optimal statistic testing the strength of correlations injected in the simulations. 
    }
    \label{fig:fp gwb}
\end{figure}

In our previous set of simulations, 
we considered an uncorrelated common process, 
but a GWB also includes the presence of interpulsar correlations described by the Hellings-Downs (HD) coefficients \cite{1983ApJ...265L..39H}. To test the effect of correlations, we created a set of simulations with white noise, intrinsic red noise, and a realistic background with HD correlations. When running our Bayesian analysis however, we once again focus only on the CURN model due to the computationally intense process of inverting a dense covariance matrix described in Sec \ref{sec: fp stat}.

We plot the \Fp\ values in Fig. \ref{fig:fp gwb} and find that the recovered values do not vary from the CURN results. This is due to the fact that the implementation of the \Fp-statistic only takes into account pulsar auto-correlations and does not consider the cross-correlations. In the bottom panel of Fig. \ref{fig:fp gwb} we also show the S/N of the injected correlations, which range from 2 to more than 12. 
We conclude that modeling the GWB as CURN does not significantly effect the \Fp\-statistic. 
To account for the presence of inter-pulsar correlations, our \Fp\-statistic could be modified to include the dense covariance matrix as discussed in Section \ref{sec: fp stat}, but this is beyond the scope of this paper.

\subsection{Continuous wave in the presence of CURN}

\begin{figure}[h]
    \centering
    \includegraphics[width=\linewidth]{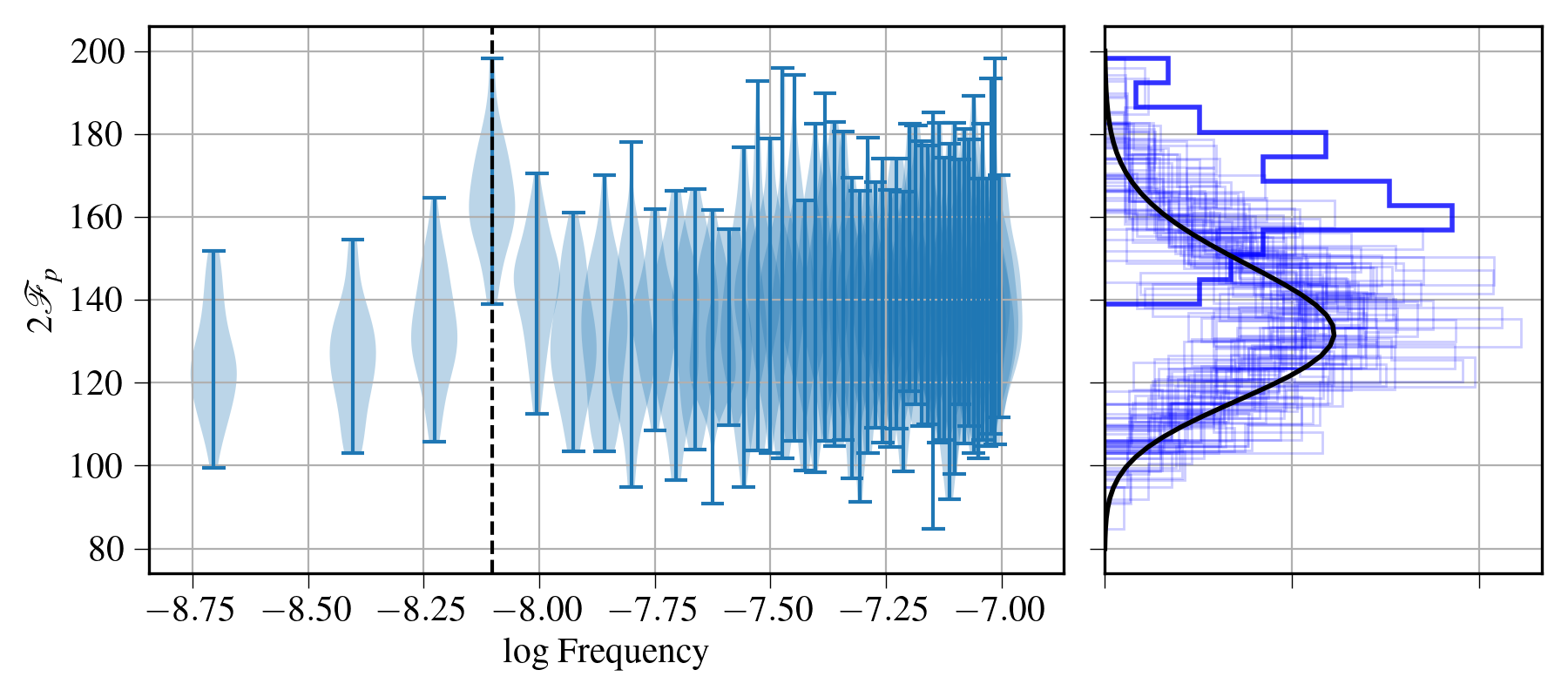}
    \includegraphics[width=\linewidth]{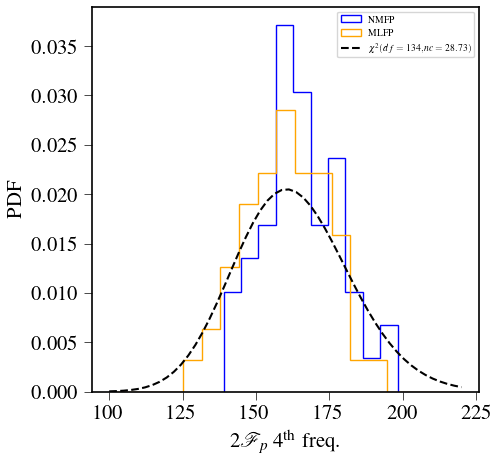}
    \caption[Top figure: Violin plots for CW simulation (Medium S/N)]{(Left) The violin plots for both the NMFP values of simulations with white noise, intrinsic red noise, a common process, and a continuous wave signal with a medium S/N. (Right) Histograms of all frequencies compared to the null distribution (black curve). The darker colored histograms indicate the fourth frequency component. Bottom figure: Histogram for the NMFP (blue) and MLFP (orange) values at the injected frequency of 8nHz compared to the expected non-central chi-squared distribution}
    \label{fig:cw fp med}
\end{figure}

Finally, we look at how well the \Fp-statistic can recover CW signals 
in the presence of white noise, intrinsic red noise, and CURN. 
We used the parameters in Table \ref{tab:cw inj params}, 
which place the source at the most sensitive sky location 
for the NANOGrav 15-year data set \citep{2023ApJ...951L..50A}. 
We kept the chirp mass fixed and 
considered two different GW frequencies, 8 nHz and 16 nHz as well as two slightly different luminosity distances,
which corresponded to signal-to-noise (S/N) ratios of 
5.4 and 9.1, respectively.

\begin{figure}[h!]
    \centering
    \includegraphics[width=\linewidth]{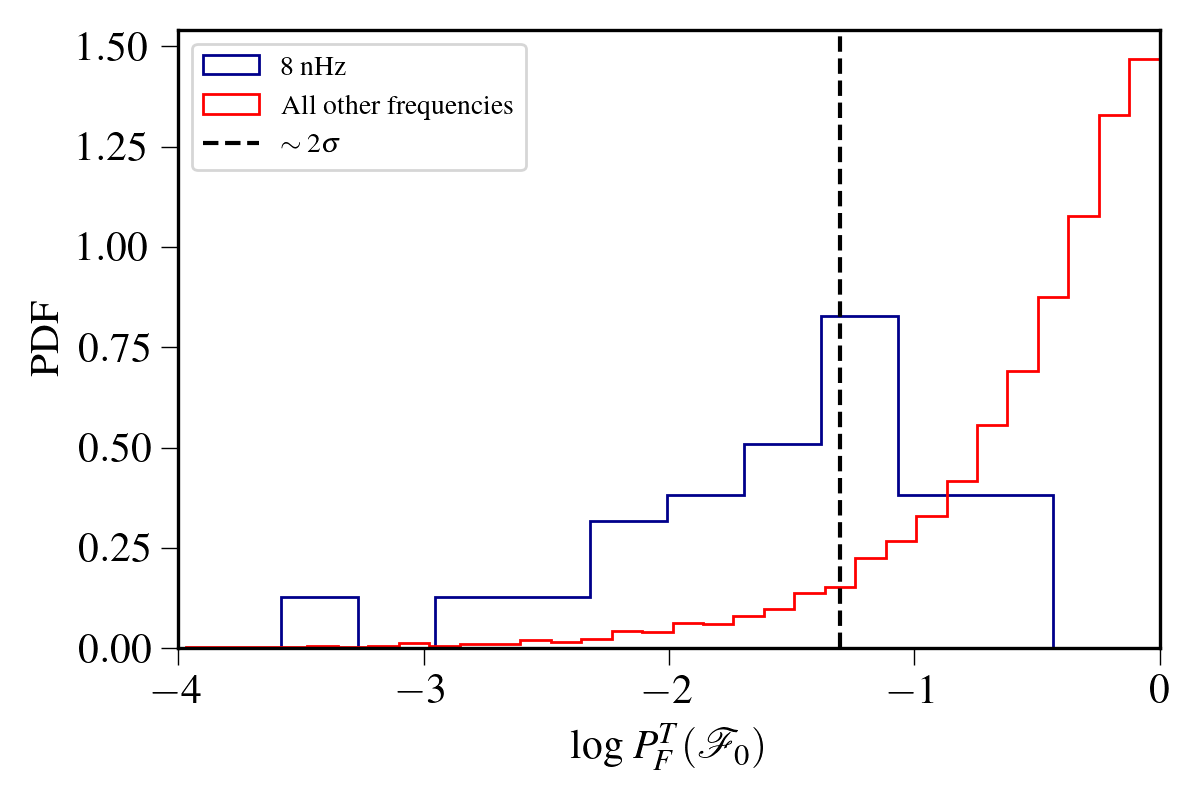}
    \caption[\shashwat{Need to replace this} FAP values for CW injection with medium S/N]{The log FAP values for the injected frequency of 8 nHz (dark blue) and all other frequencies (red). The dashed-line indicates a p-value of 0.05, corresponding to $\sim 2 \sigma$. At the injected frequency, 56\% of NMFP values have an FAP value less than 0.05. None of the simulations fall below the 11yr threshold of $10^{-4}$.}
    \label{fig:fap cw med}
\end{figure}

\subsubsection{8 nHz injection (S/N $=5.4$)}

Figure~\ref{fig:cw fp med} shows the violin plots of $2\mathcal{F}_p$ for a CW injection at 8 nHz. 
We see that at the fourth frequency bin, 
which corresponds to the frequency of the injected CW, 
2\Fp\ does not follow a chi-squared distribution with $\rho^2=0$, 
but it does at the other frequencies. In the bottom panel of Fig. \ref{fig:cw fp med} we show the histogram of the NMFP and MLFP values at the injected frequency of 8nHz. We show that the recovered values follow the expected chi squared distribution with a non centrality of 28.73. Figure~\ref{fig:fap cw med} shows a histogram of the p-values for $2\Fp$ 
at 8 nHz. We find that at 8 nHz  56\% of simulations have $p < 0.05$. 
In contrast, at every other frequency, 
we find $p < 0.05$ in 9.33\% of simulations. 
From this we conclude that, 
although the \Fp-statistic values at 8 nHz 
follow a different distribution than at the other frequencies, 
the significance is too low to be considered a detection.

\subsubsection{16 nHz injection (S/N $=9.1$)}

\begin{figure}[h!]
    \centering
    \includegraphics[width=\linewidth]{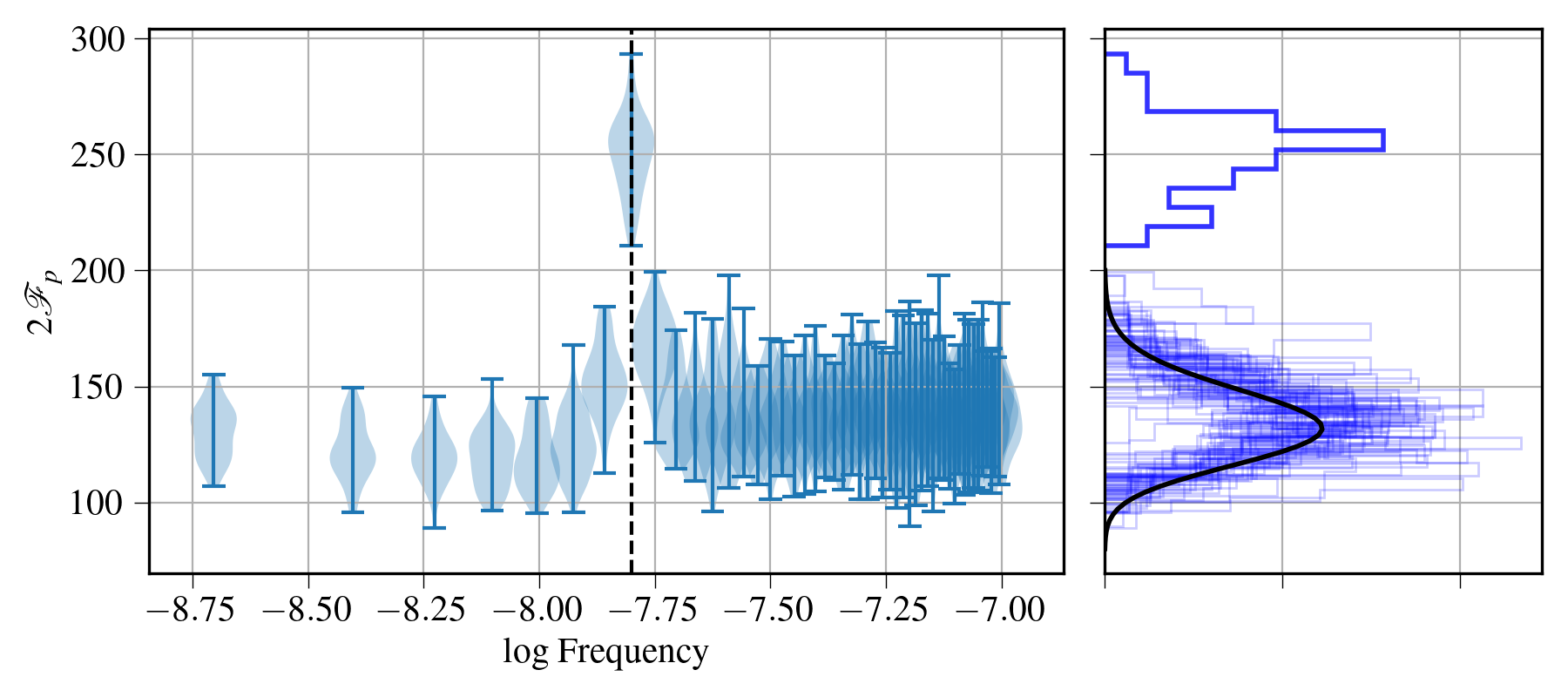} \\
    \includegraphics[width=0.9\columnwidth]{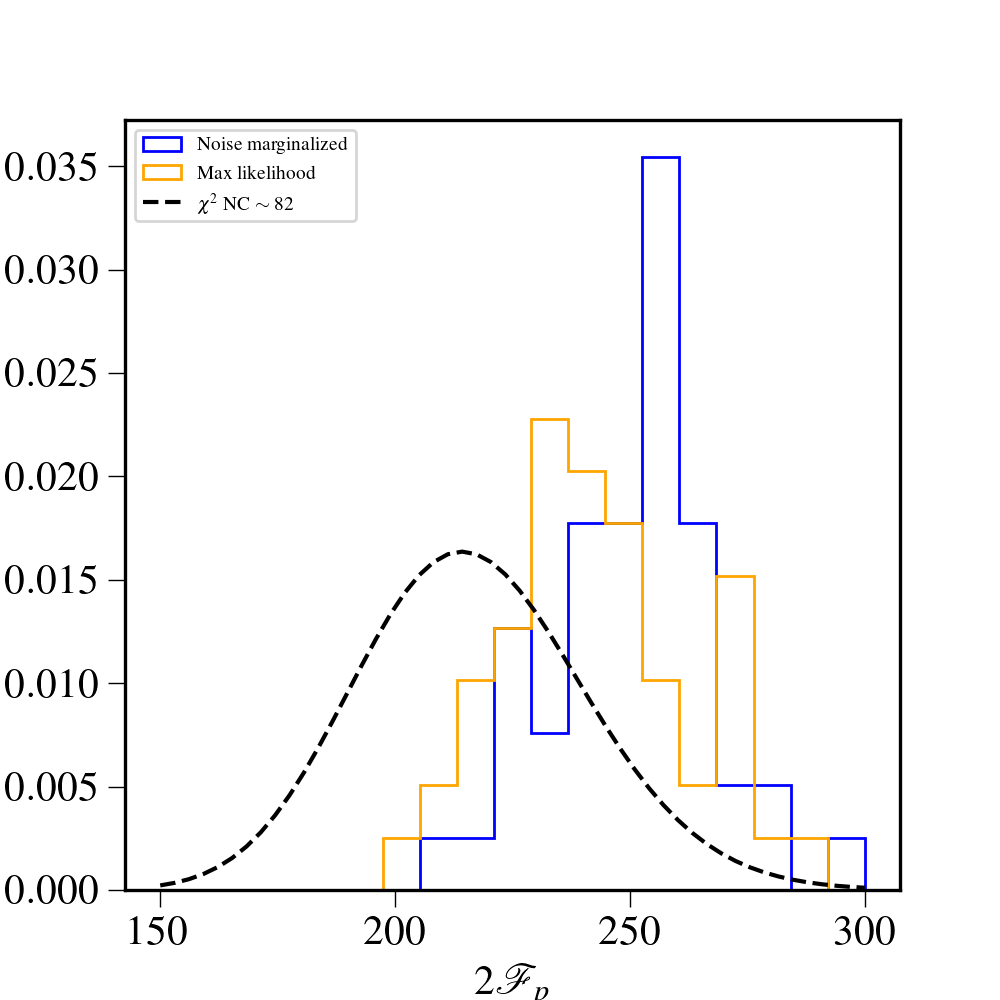}
    \caption[Violin plots for high S/N CW simulations]{Top figure: (Left) The violin plots for the mean NMFP values of 50 simulations with white noise, intrinsic red noise, a common process, and a high S/N continuous wave as a function of frequency. (Right) The histograms for the $2\mathcal{F}_p$ values as compared to the expected central chi-squared distribution with 134 degrees of freedom. Bottom figure: The histograms for the \Fp statistic at 16 nHz compared to the expected chi squared distribution with non-centrality of 82.}
    \label{fig:cw fp}
\end{figure}

Figure~\ref{fig:cw fp} shows the violin plots of the $2\mathcal{F}_p$ values. We see that at the eighth frequency, which corresponds to the frequency of the CW injection, 
$2\Fp$ does not follow a chi-squared distribution 
with non-centrality parameter of zero. 
There is also a shift in the distribution of $2\Fp$ 
at the frequencies near the CW injection, 
especially the ninth frequency ($\sim18$ nHz). 
At all other frequencies however, the distributions follows the expected central chi-squared distribution. 
In the bottom panel of Fig. \ref{fig:cw fp} we see that the recovered NMFP and MLFP values at the injected frequency follow a distribution with a non-centrality that is $\sim 1.2 \times$ the expected non-centrality.

\begin{figure}[t!]
    \centering
    \includegraphics[width=\linewidth]{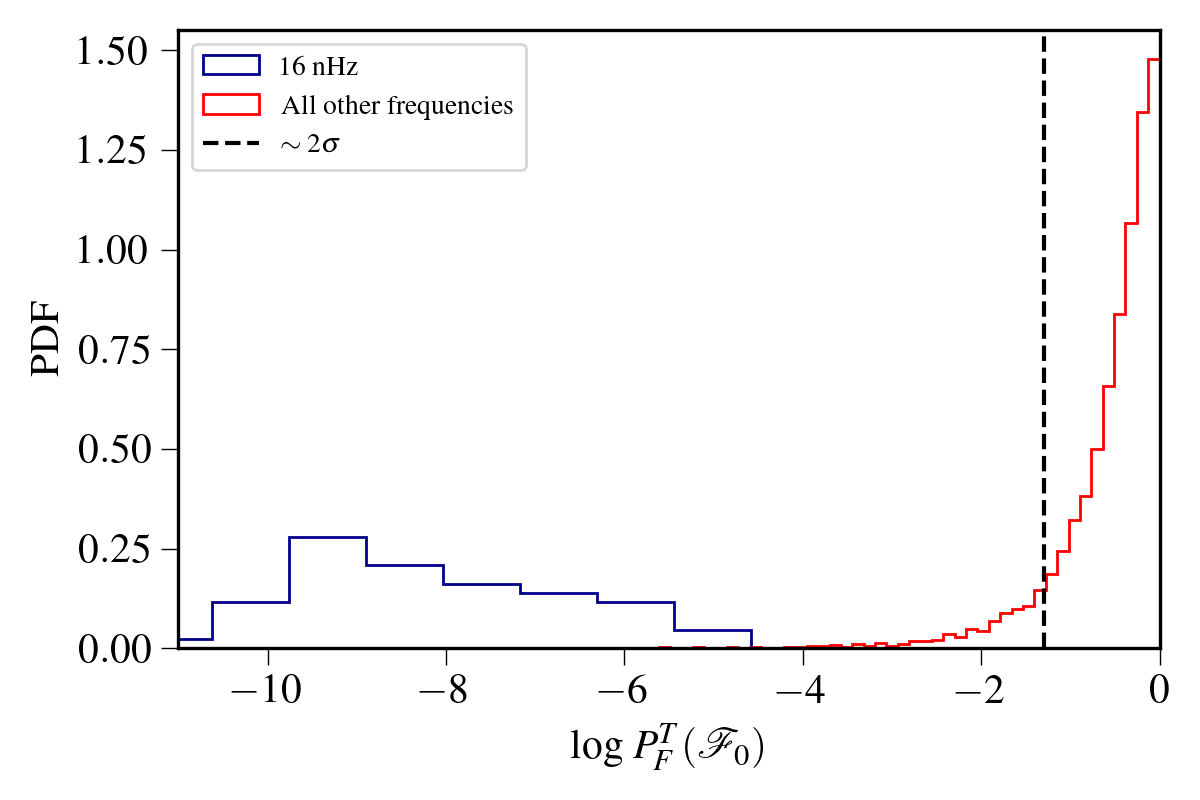}
    \caption[FAP values for high S/N CW simulations]{Histogram of log FAP values for the injected frequency at 16 nHz (dark blue) and all other frequencies (light blue). The dashed line indicates a p-value of 0.05, corresponding to $\sim 2 \sigma$. All simulations have $p < 10^{-4}$ at the injected frequency.}
    \label{fig:fap cw}
\end{figure}

In Fig. \ref{fig:fap cw}, we show the FAP values calculated from the mean NMFP values for 50 simulations at 16 nHz. 
We find that all of the simulations have $p < 0.05$, and 88\% of simulations have $p < 10^{-4}$ at the injected frequency of 16 nHz. 
In contrast, across the other frequencies 
we find $p < 0.05$ in 9.3\% of simulations, 
and $p < 10^{-4}$ in 0.3\% of simulations. 
From this we conclude that the \Fp-statistic is able to detect a CW at the injected frequency of 16 nHz in the majority of simulations.

\section{Results: NANOGrav 12.5-year data set} \label{sec: 12.5yr results}

In this section we present the $\mathcal{F}_p$ values of the NANOGrav 12.5 yr dataset of 45 pulsars with noise marginalization. 
In \cite{2023ApJ...951L..28A} we presented only the Bayesian analysis results, because we had not previously used the \Fp-statistic when a common process was also present. We use intrinsic red noise models with 30 frequency components, and a common uncorrelated red noise model of 5 frequency components as was used for the analysis of the 12.5 yr dataset in \cite{2020ApJ...905L..34A}. The real dataset also includes white noise processes which are correlated among different epochs (ECORR) which we did not include in our simulations. Since this white noise is correlated at different observing epochs it is possible that it results in a slightly different distribution in $\mathcal{F}_p$ values than the null distribution.

\begin{figure}[h]
    \centering
    \includegraphics[width=\linewidth]{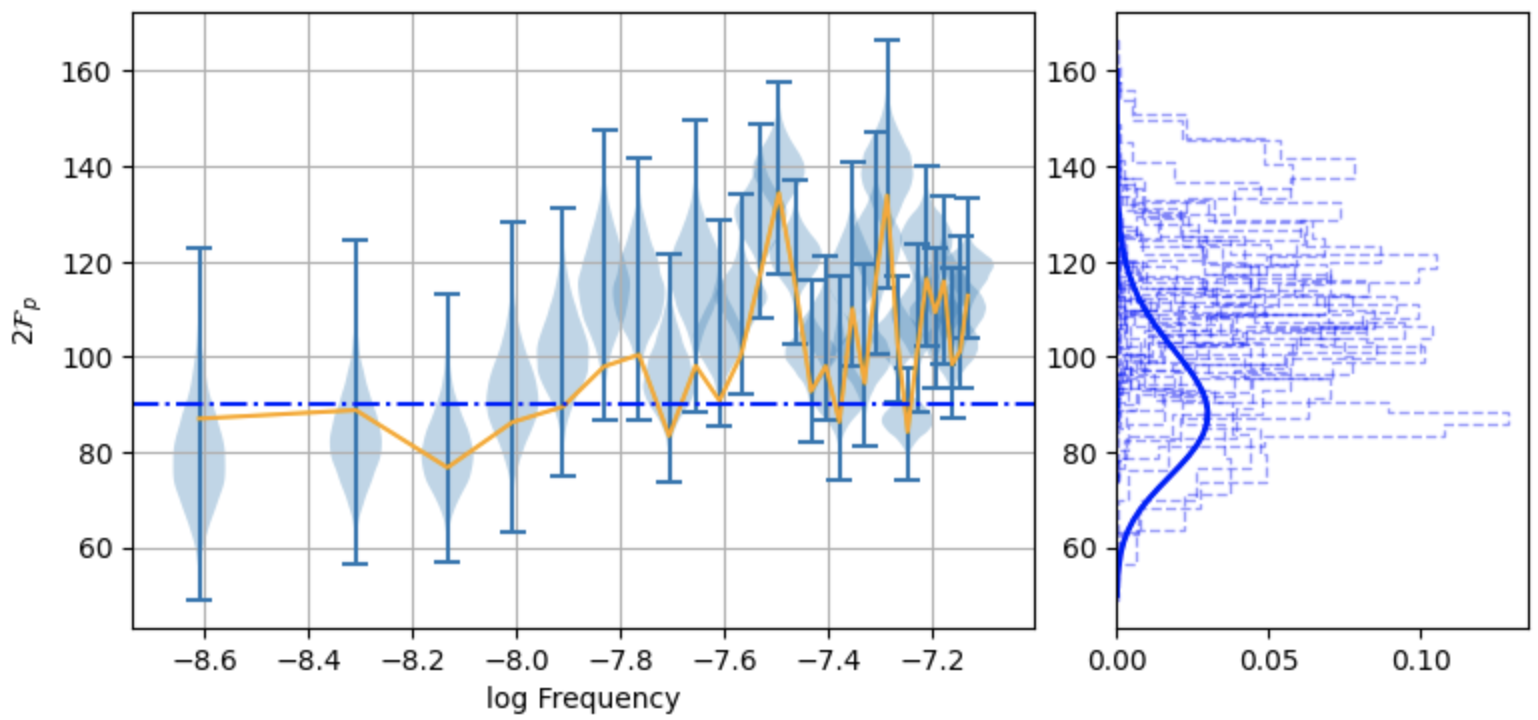}
    \caption{The $2\mathcal{F}_p$ values for the 12.5 yr dataset. On the left panel, the blue violin plots are the NMFP values for 20000 red noise parameters. The orange line is the noise maximized $\mathcal{F}_p$ values. The dot-dashed blue line is the expected mean for a chi squared distribution with 90 degrees of freedom. In the right panel, the dark blue histogram is the null chi-squared distribution with 90 degrees of freedom and zero non-centrality.}
    \label{fig:12p5yr}
\end{figure}

In Fig. \ref{fig:12p5yr} we show the violin plot of the noise-marginalized $\mathcal{F}_p$ values for the 12.5 yr dataset. We see that the first few frequencies are within the expected range of the null hypothesis. After the first five frequency components, where we stop modeling the common process, we get higher $\mathcal{F}_p$ values. At $f_\mathrm{gw} = 1 \;\mathrm{yr}^{-1} = 31.9$ nHz, we recover higher values due to the insensitivity of PTAs to GW's at this frequency owing to the Earth's orbital period. We also see higher $\mathcal{F}_p$ values at $\sim 51.5 \mathrm{nHz}$.

\begin{figure}
    \centering
    \includegraphics[width=0.75\linewidth]{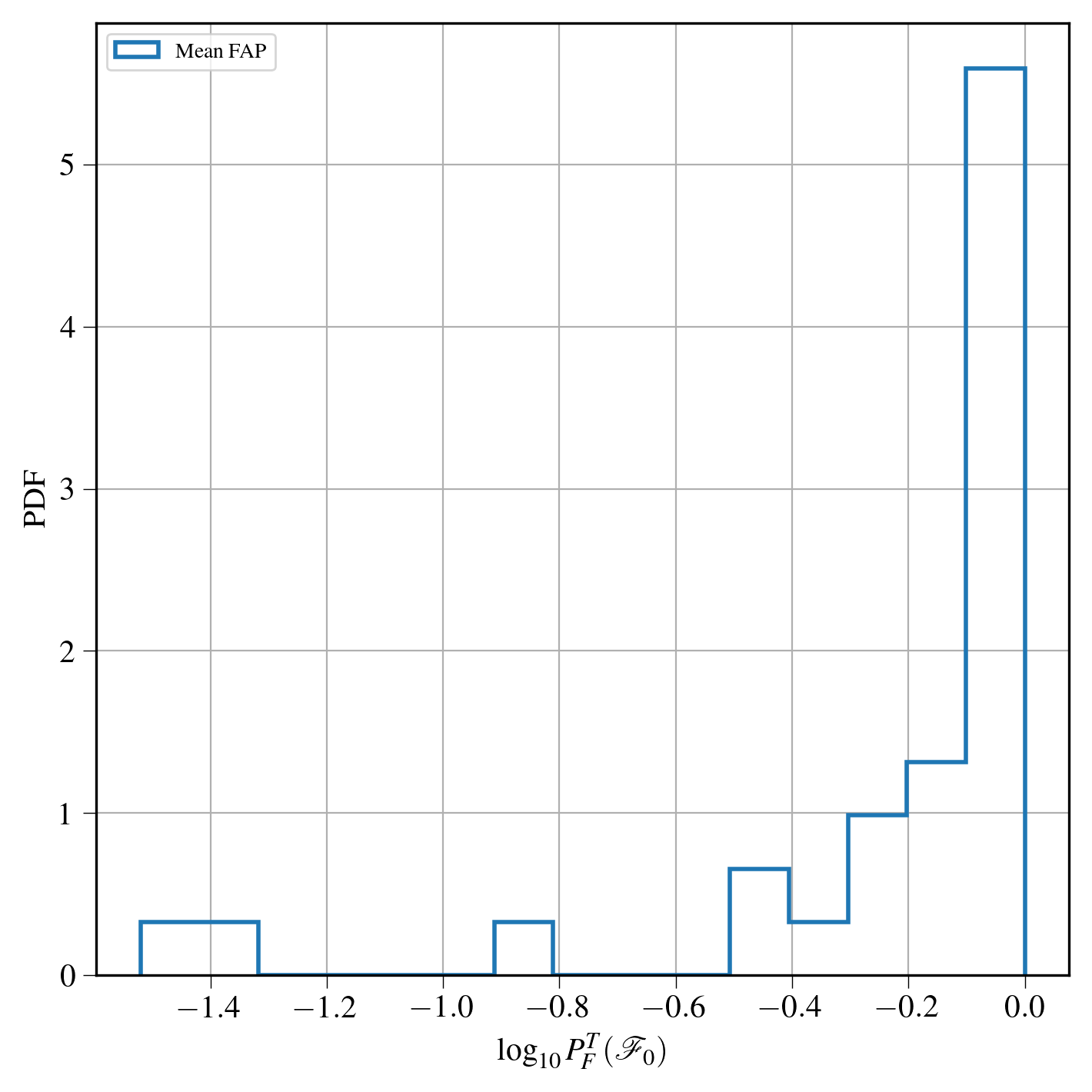}
    \includegraphics[width=0.75\linewidth]{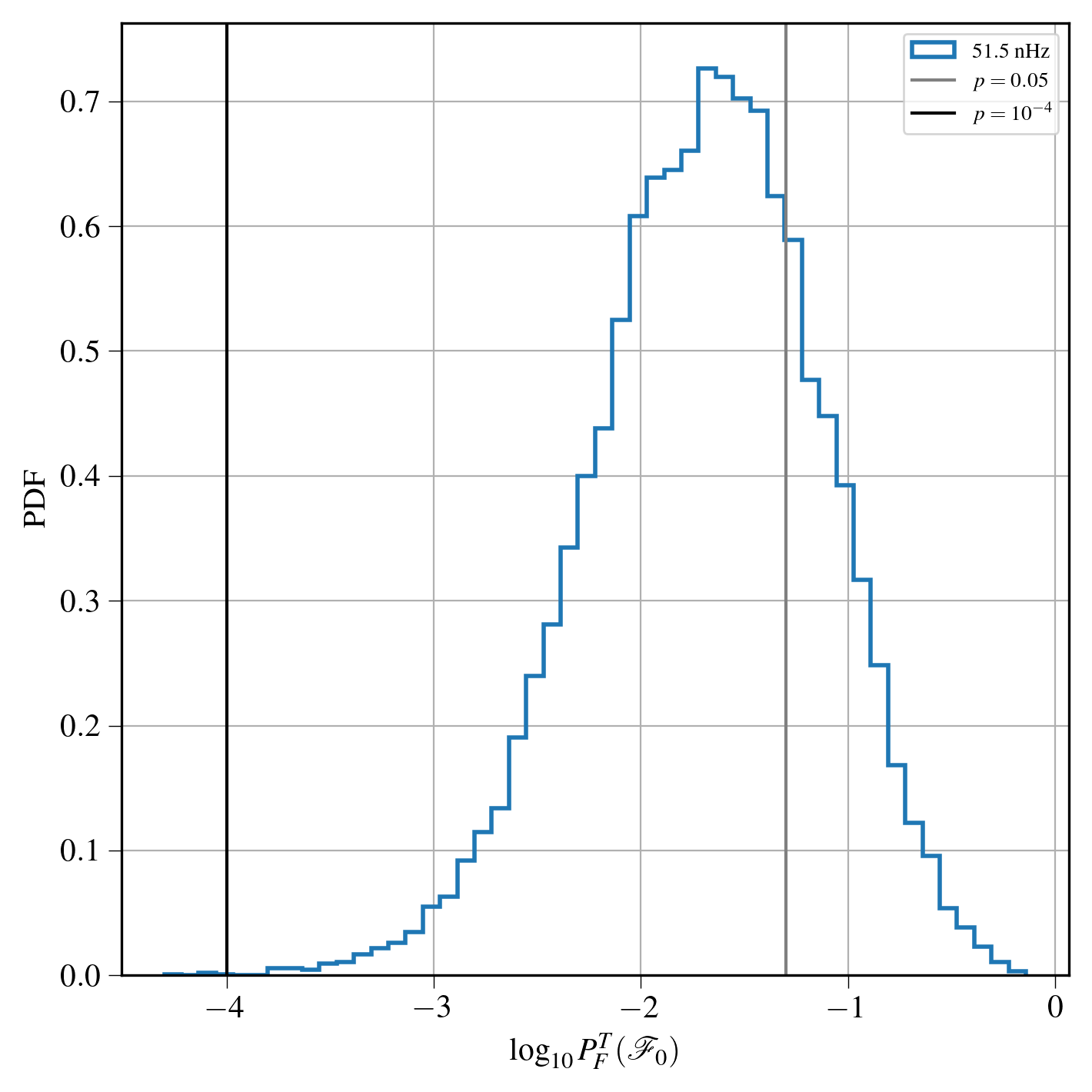}
    \caption{(TOP) Mean FAP values, marginalized over noise posteriors, for each frequency in the 12.5yr data set. None of the frequencies have a mean FAP below $10^{-4}$, and the lowest mean FAP value occurs at 51.5 nHz, with a value of 0.04. (BOTTOM) The full distribution of FAP values at 51.5 nHz over noise parameters. We see that more than half of the values fall below the 0.05 threshold, and only 0.02\% fall below the threshold of $10^{-4}$. Therefore, we conclude that there is no evidence of a CW in the 12.5yr data set. This result is consistent with the Bayesian results published in \cite{2023ApJ...951L..28A}.}
    \label{fig:12p5yr fap}
\end{figure}

In Fig. \ref{fig:12p5yr fap} we plot the mean FAP values of the calculated NMFP values for the 12.5 yr dataset.
In the bottom panel of Fig.~\ref{fig:12p5yr fap}, we show the distribution of FAP values at 51.5 nHz over noise parameters. At this frequency, the mean FAP value is 0.04, and a small fraction of noise parameters (0.02\%) of FAP values have $p<10^{-4}$. Comparing to the simulations, a mean FAP value of 0.04 is consistent with both the medium S/N continuous wave simulations and the no CW simulations. 
Taking into account the trials factor from searching over 
30 frequencies, we conclude that these results are consistent 
with only noise being present in the data: 
the $p$-value with the Bonferroni correction is $p = 0.04 \times 30 \sim 1$.
We note that Bayesian searches for CWs have found 
that unmodeled noise in individual pulsars 
has a significant effect on Bayes factors \cite{NG_11yr_cw,2023ApJ...951L..28A,2023ApJ...951L..50A}. 
In this analysis, we have not incorporated 
any advance noise modeling: doing so could result in 
lower \Fp-statistic values and higher FAP values 
since unmodeled noise can be confused for a CW.

\section{Conclusions} \label{sec: conclusions}

In this paper, 
we used simulations to study the performance of the \Fp-statistic 
in the presence of pulsar red noise and common red noise consistent with a GWB. 
we introduced noise-marginalization to the \Fp -statistic and compared it to the previously used \Fp values calculated from the maximum likelihood noise parameters. We found that the NMFP and MLFP give similar results to each other and are able to recover accurate values in the presence of white noise and intrinsic pulsar red noise. 

We were also able to recover the expected \Fp values even in the presence of white noise, intrinsic red noise, and a common process, with or without correlations. This is due to the fact that the \Fp -statistic is only concerned with pulsar auto-correlations, and not cross-correlations. In previous analyses of the \Fp -statistic, models only included white noise, or white noise and a common uncorrelated process. We have shown that the NMFP can accurately retrieve values for more datasets with more realistic noise descriptions. We also showed that using the maximum likelihood GWB parameters is able to recover comparable \Fp values to marginalizing over GWB parameters.

We showed that both the NMFP and MLFP were comparable even in the presence of a continuous wave with medium or high S/N. We found that for the medium S/N case, the recovered \Fp\ values followed a $\chi^2$ distribution with the expected non-centrality. In the case of the high S/N simulations, we found that both the MLFP and NMFP recovered \Fp\ values that followed a $\chi^2$ distribution with a larger non-centrality than expected, but were still consistent with a detection.

For both sets of continuous wave simulations, we measured the significance of the detection using the FAP values, with thresholds of 0.05 and $10^{-4}$. We showed that in the case of the medium S/N more than half of simulations recovered a value less than 0.5 at the injected frequency. For the high S/N case, all the values at the injected frequency fell below the 0.05 threshold and nearly 90\% of them fell below the $10^{-4}$ threshold.

We also reanalyzed the NANOGrav 12.5yr dataset using the NMFP and compared it to the MLFP values. We found no evidence of a CW in our analysis, consistent with the previously published Bayesian results. The lowest mean FAP value was 0.04 at a frequency of 51.5 nHz; accounting for the trials factor gives $p\sim1$.

We have shown that the \Fp-statistic is a robust tool for detecting CWs in the presence of pulsar red noise and a GWB. 
We get similar results whether we marginalize over noise posteriors or use the maximum-likelihood values.  This in contrast to frequentist searches for the GWB using the optimal statistic \cite{2018PhRvD..98d4003V}, where the common process is highly covariant with intrinsic pulsar red noise.
Frequentist searches with the \Fp-statistic complement Bayesian searches. The \Fp-statistic uses a simplified signal model, but is significantly faster to compute compared to performing a Bayesian analysis. 
With \fastfp\ the calculations can be accomplished in a matter of minutes using GPU acceleration, 
making it a useful tool for performing 
rapid searches for CWs 
across the PTA spectrum.

\section*{Acknowledgments}

We would like to thank Rand Burnette, Bjorn Larson, and Chiara Mingarelli for their comments and helpful insights. SCS and SJV are supported by NSF award number 2309246. GEF is supported by National Aeronautics and Space Administration (NASA) Future Investigators in NASA Earth and Space Science and Technology Grant No. 80NSSC22K1591. CAW acknowledges support from CIERA, the Adler Planetarium, and the Brinson Foundation through a CIERA-Adler postdoctoral fellowship. The authors are members of the NANOGrav collaboration, which is supported by NSF Physics Frontiers Center award number 2020265.


\appendix
\section{Injected red noise parameters}

In Table \ref{tab:rn params fp} we present the intrinsic red noise parameters injected in our datasets
\begin{table}[t!]
    \caption[Injected red noise parameters for $\mathcal{F}_p$ statistic]{The injected intrinsic red noise parameters (spectral index and log amplitude) for all the simulations, as well as the common process}
    \begin{center}
    \begin{adjustbox}{max width=9cm}
    \begin{tabular}{|l|c|c||l|c|c|}
    \hline
        $\textbf{Pulsar}$ & $\boldsymbol{\gamma}$ & $\boldsymbol{\log_{10}A}$ & $\textbf{Pulsar}$ & $\boldsymbol{\gamma}$ & $\boldsymbol{\log_{10}A}$ \\
    \hline  \hline
        B1855+09 & 4.6094 & $-$14.1708 & J1730$-$2304 & 2.108 & $-$18.859\\
    \hline
        B1937+21 & 4.0926 & $-$13.633 & J1738+0333 & 4.026 & $-$14.210\\
    \hline
        B1953+29 & 1.894 & $-$12.907 & J1741+1351 & 3.629 & $-$19.695\\
    \hline
        J0023+0923 & 0.3826 & $-$14.182 & J1744$-$1134 & 5.677 & $-$18.795\\
    \hline
        J0030+0451 &  5.488 & $-$15.012 & J1745+1017 & 2.445 & $-$11.921\\
    \hline
        J0340+4130 & 3.323 & $-$15.860 & J1747$-$4036 & 3.033 & $-$12.791\\
    \hline
        J0406+3039 & 3.769 & $-$16.043 & J1751$-$2857 & 2.111 & $-$18.693\\
    \hline
        J0437$-$4715 & 0.9381 & $-$18.160 & J1802$-$2124 & 0.604 & $-$12.123\\
    \hline
        J0509+0856 & 0.231 & $-$12.963 & J1811$-$2405 & 2.093 & $-$19.812\\
    \hline
        J0557+1551 & 6.183 & $-$19.043 & J1832$-$0836 & 0.587 & $-$15.831\\
    \hline
        J0605+3757 & 0.075 & $-$17.611 & J1843$-$1113 & 2.308 & $-$16.670\\
    \hline
        J0610$-$2100 & 2.549 & $-$12.688 & J1853+1303 & 0.428 & $-$13.270\\
    \hline
        J0613$-$0200 & 2.977 & $-$13.667 & J1903+0327 & 1.376 & $-$12.162\\
    \hline
        J0636+5128 & 3.911 & $-$15.842 & J1909$-$3744 & 2.215 & $-$17.9584\\
    \hline
        J0645+5158 & 1.677 & $-$13.538 & J1910+1256 & 0.689 & $-$15.539\\
    \hline
        J0709+0458 & 2.537 & $-$12.857 & J1911+1347 & 4.680 & $-$15.506\\
    \hline
        J0740+6620 &  0.030 & $-$14.636 & J1918$-$0642 & 3.885 & $-$17.894\\
    \hline
        J0931$-$1902 & 3.444 & $-$15.315 & J1923+2515 & 0.996 & $-$18.762\\
    \hline
        J1012+5307 & 0.498 & $-$12.565 & J1944+0907 & 1.074 & $-$13.066\\
    \hline
        J1012$-$4235 & 2.528 & $-$13.671 & J1946+3417 & 0.537 & $-$12.441\\
    \hline
        J1022+1001 & 1.015 & $-$12.106 & J2010$-$1323 & 2.325 & $-$15.567\\
    \hline
        J1024$-$0719 & 1.134 & $-$15.699 & J2017+0603 & 3.549 & $-$18.739\\
    \hline
        J1125+7819 & 5.193 & $-$17.020 & J2033+1734 & 5.221 & $-$15.429\\
    \hline
        J1312+0051 & 0.694 & $-$12.776 & J2043+1711 & 1.488 & $-$17.604\\
    \hline
        J1453+1902 & 2.508 & $-$19.971 & J2124$-$3358 & 1.308 & $-$19.499\\
    \hline
        J1455$-$3330 & 3.688 & $-$16.338 & J2145$-$0750 & 0.264 & $-$12.814\\
    \hline
        J1600$-$3053 & 3.075 & $-$15.765 & J2214+3000 & 4.795 & $-$16.653\\
    \hline
        J1614$-$2230 & 6.615 & $-$18.599 & J2229+2643 & 6.826 & $-$16.640\\
    \hline
        J1630+3734 & 3.665 & $-$14.084 & J2234+0611 & 6.628 & $-$15.630\\
    \hline
        J1640+2224 & 4.970 & $-$16.783 & J2234+0944 & 1.476 & $-$15.908\\
    \hline
        J1643$-$1224 & 1.240 & $-$12.296 & J2302+4442 & 1.403 & $-$15.363\\
    \hline
        J1705$-$1903 & 0.463 & $-$12.073 & J2317+1439 & 6.896 & $-$17.790\\
    \hline
        J1713+0747 & 4.139 & $-$16.175 & J2322+2057 & 2.020 & $-$14.796\\
    \hline
        J1719$-$1438 & 4.186 & $-$16.637 & CURN & 4.333 & $-$14.699\\
    \hline
    
    \end{tabular}
    \end{adjustbox}
    \end{center}
    
\label{tab:rn params fp}
\end{table}

\section{CW parameters}

In Table \ref{tab:cw inj params} we show the parameters governing the CW injections in both cases. The sky position is the most sensitive at 6 nHz from the NANOGrav 15yr dataset \cite{2023ApJ...951L..50A}

\begin{table}
    \caption[CW parameters]{The injected parameters for the deterministic individual source}
    \begin{center}
    \begin{adjustbox}{max width=12cm}
    \begin{tabular}{|l|c|}
    \hline \hline
        Parameters & Injected value \\
    \hline
        $\mathcal{M}_c$ & $10^9 ~M_\odot$ \\
    \hline
        $f_{\mathrm{cw}}$ & $8, 16~ \mathrm{nHz}$ \\
    \hline
        $d_L$ & $84.3 ~\mathrm{Mpc}$, $87.76 ~\mathrm{Mpc}$ \\
    \hline
        $\theta$ & 0 \\
    \hline
        $\phi$ & $7\pi/4$ \\
    \hline
        $\Phi_0$ & 0 \\
    \hline
        $\iota$ & 0 \\
    \hline
        $\psi$ & 0 \\
    \hline
        SNR &  5.36, 9.1 \\
    \hline
    \end{tabular}
    \end{adjustbox}
    \end{center}
    
\label{tab:cw inj params}
\end{table}

\section{\Fp recovery} \label{p-p plots}

\begin{figure*}[h!]
    \centering
    \includegraphics[width=0.6\linewidth]{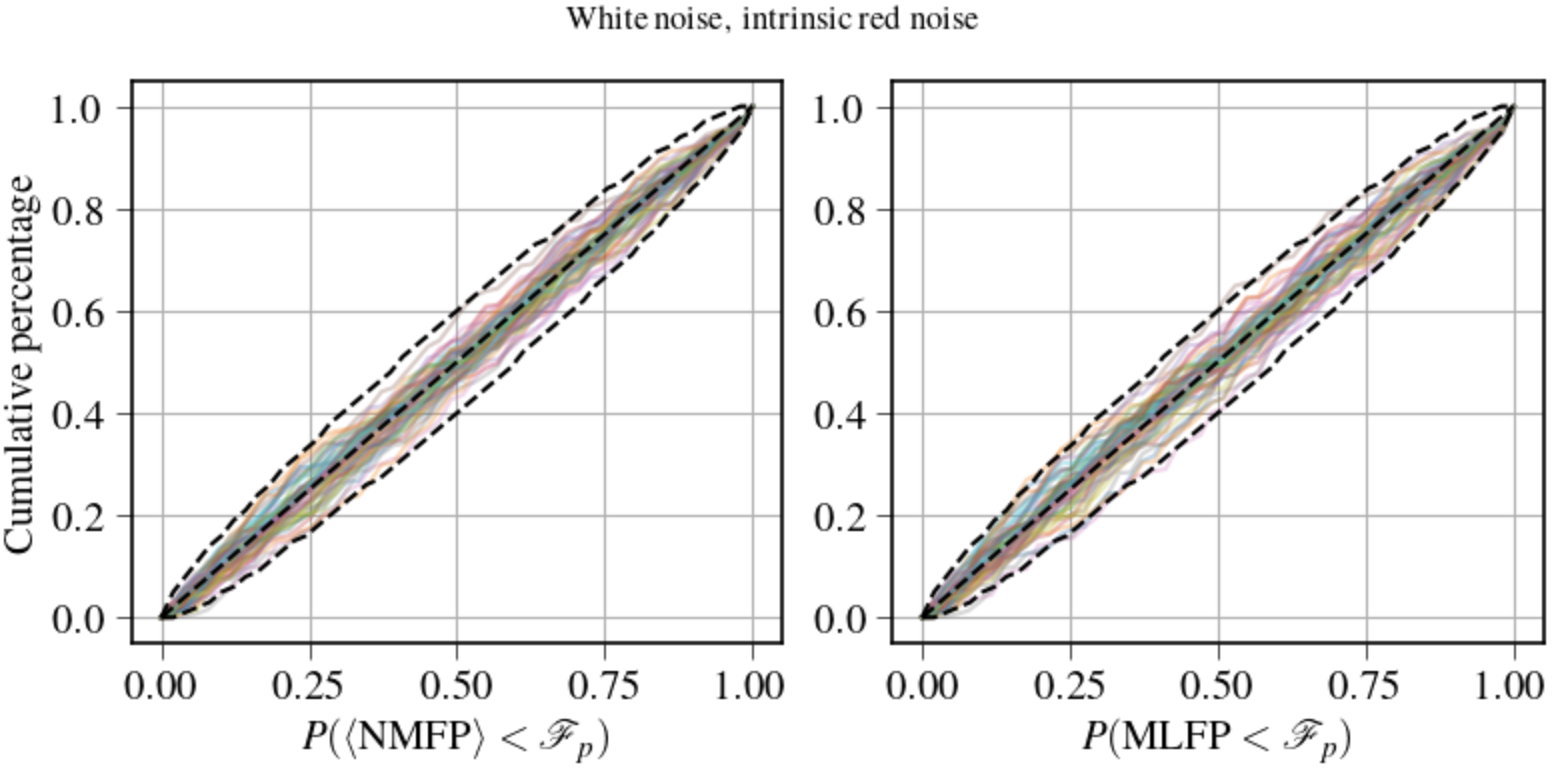} \includegraphics[width=0.6\linewidth]{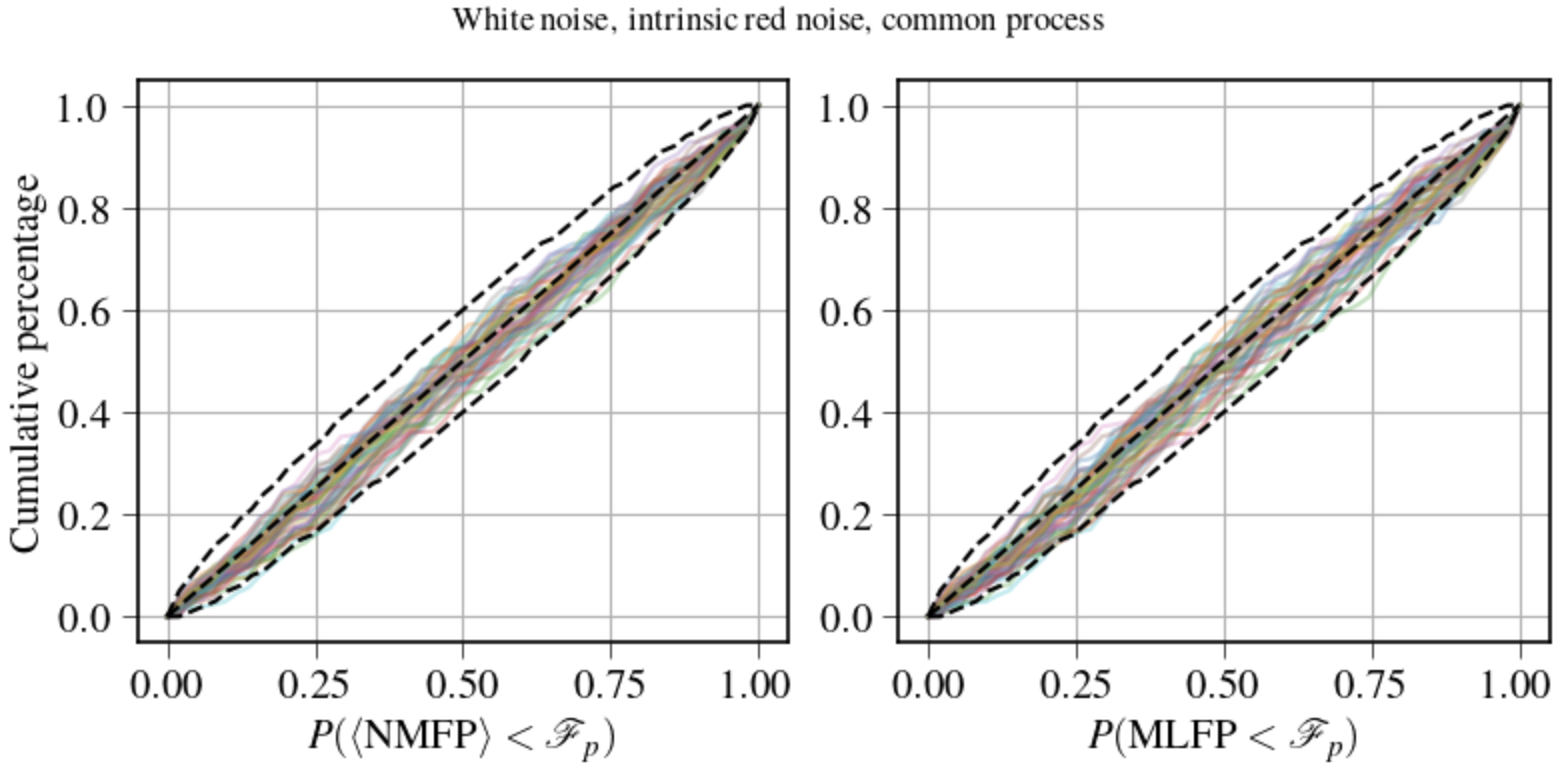}
    \includegraphics[width=0.6\linewidth]{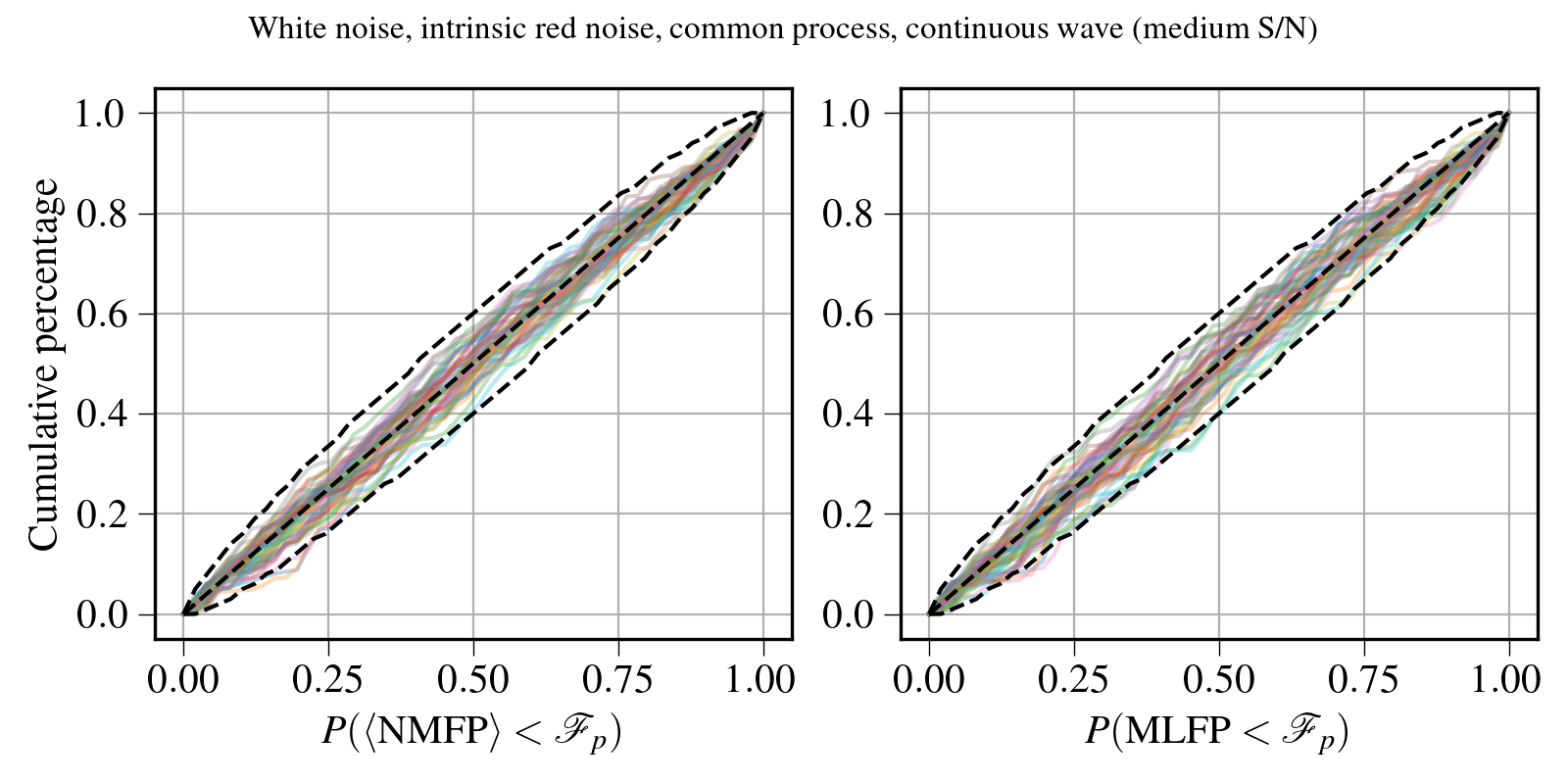}
    \includegraphics[width=0.6\linewidth]{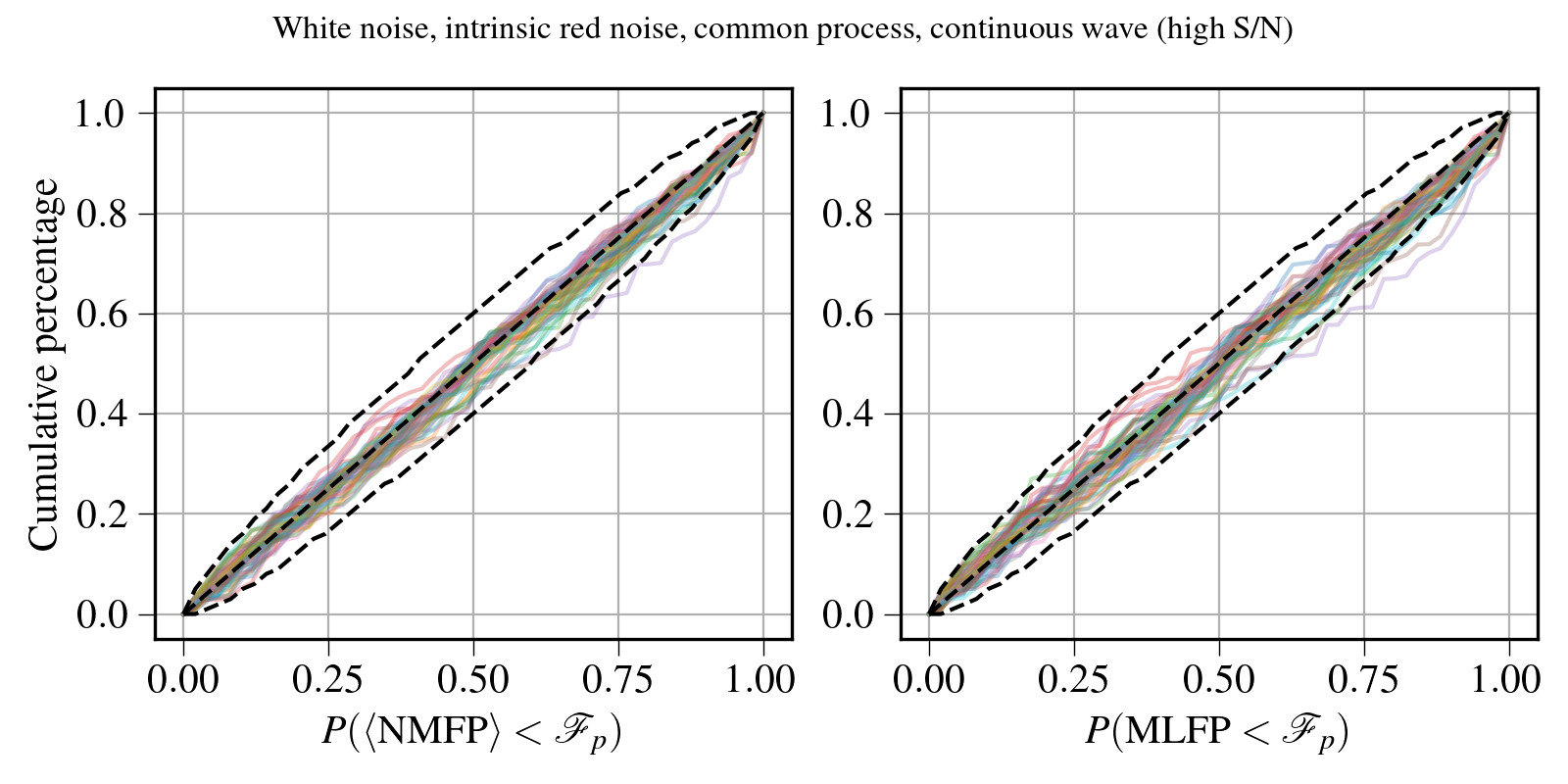}
    \caption{$p-p$ diagrams for the four different types of simulations. The left panels show the recovery of the mean NMFP values, and the right panels show the recovery of the MLFP values. The dashed lines indicate a 95\% confidence interval.}
    \label{fig:p-p diagrams}
\end{figure*}

We show the recovery of the \Fp values for each frequency across all simulations using $p-p$ diagrams in Fig. \ref{fig:p-p diagrams}. We show the cumulative percentage of the \Fp values compared to the expected distributions across the 50 modeled frequencies.

\bibliographystyle{apsrev}
\bibliography{authors}

\end{document}